\documentclass[twocolumn]{aastex63}
\usepackage{color} 
\usepackage{tabularx} 
\usepackage{epsfig}
\usepackage{amsmath} 
\usepackage{amssymb} 
\usepackage{graphicx}
\usepackage{cprotect}
\usepackage{url}
\graphicspath{{./}{figures/}}
\definecolor{purp}{HTML}{8904B1}

\newcommand{\chandra}{\emph{Chandra}\ }
\newcommand{\picm}{\rho_{\text{\tiny ICM}}}
\newcommand\fe[3]{$#1^{+#2}_{-#3}$}

\shortauthors{Watson et al.}

\begin{document}

\title{\emph{CHANDRA} X-RAY OBSERVATIONS OF ABELL 119: \\ COLD FRONTS AND A SHOCK IN AN EVOLVED OFF-AXIS MERGER}

\author[0000-0001-8456-4142]{Courtney B. Watson}
\affil{Institute for Astrophysical Research and Department of Astronomy, Boston University, Boston, MA 02215, USA}
\email{cbwatson@bu.edu}

\author[0000-0002-0485-6573]{Elizabeth L. Blanton}
\affil{Institute for Astrophysical Research and Department of Astronomy, Boston University, Boston, MA 02215, USA}

\author[0000-0002-3984-4337]{Scott W. Randall}
\affil{Harvard Smithsonian Center for Astrophysics, 60 Garden Street, Cambridge, MA 02138, USA}

\author[0000-0003-0167-0981]{Craig L. Sarazin}
\affil{Department of Astronomy, University of Virginia, P.O. Box 400325, Charlottesville, VA 22904-4325, USA}

\author[0000-0002-5222-1337]{Arnab Sarkar}
\affil{Kavli Institute for Astrophysics and Space Research, Massachusetts Institute of Technology, 77 Massachusetts Ave, Cambridge, MA 02139, USA}

\author[0000-0003-3175-2347]{John A. ZuHone}
\affil{Harvard Smithsonian Center for Astrophysics, 60 Garden Street, Cambridge, MA 02138, USA}

\author[0000-0003-3646-3472]{E. M. Douglass}
\affil{Farmingdale State College—SUNY, 2350 Broadhollow Rd., Farmingdale, NY 11735, USA}

\accepted{08/07/2023}
\submitjournal{The Astrophysical Journal}

\begin{abstract}
    We present \chandra X-ray observations of the dynamically complex galaxy cluster Abell 119 ($z = 0.044$). A119 is host to two NAT radio sources (0053-015 \& 0053-016) whose tails are oriented parallel to each other despite orthogonally oriented jet axes. Imaging and spectral analysis reveal X-ray emission elongated along the NE-SW axis along with the presence of complex structures, including surface brightness discontinuities, which suggest possible merger activity along this axis. From radial profiles of the X-ray surface brightness, temperature, pressure, and density, we identify two surface brightness edges which are found to be cold fronts, possibly associated with large-scale sloshing of ICM gas. We also identify a brightness edge to the south which is found to be a shock front with Mach number $M = 1.21 \pm 0.11$, consistent with a merger shock. In addition, previous optical studies show alignment of optical substructures along the north-south direction. The elongated X-ray emission, orientations of the NAT tails, and alignment of optical substructure all suggest recent or on-going merger activity in the NE-SW direction.
\end{abstract}

\section{Introduction}

Galaxy clusters are the largest gravitationally bound objects in the universe, containing hundreds to thousands of galaxies. In the current model of large-scale structure formation, galaxy clusters grow through mergers of smaller systems in a process called hierarchical structure formation \citep{West1995}. Galaxy cluster mergers are some of the most energetic events in the universe. During a merging event, this energy can be dispersed throughout the gas of the intracluster medium (ICM), leaving observable impacts on the ICM structure. 

In merging systems, bulk motions of ICM gas can give rise to density discontinuities observed as edges in the surface brightness in X-ray observations of the hot ICM. Surface brightness edges typically seen in merging systems are indicative of either shocks or cold fronts. 

The formation of shock fronts resulting from cluster mergers should be fairly common and has been seen in numerous simulations \citep{Schindler1993, Roettiger1993, Burns1998}. However, while evidence of shocks has been seen, i.e., regions of ICM showing shock-heating \citep{Markevitch2001, Markevitch2003, Kempner2004, Govoni2004}, observation of shock fronts, with both a sharp edge in gas density and a distinct jump in temperature, have only been seen in a handful of clusters. The lack of direct observations of shock fronts is mostly because we have to observe the merger at the right time (i.e. before the shock has propagated out to the low-brightness outskirts) and from the right orientation (i.e the tangent plane of the shock front edge must be close to perpendicular to the plane of the sky). Despite the challenges of observing shock fronts, when we are lucky enough to see/study them, they can provide important insight into the merger history and dynamics of the cluster.

Cold fronts (or contact discontinuities), which are more easily observed than shock fronts \citep{Markevitch2012}, have proven to be invaluable tools in understanding the physical properties of the ICM \citep{Markevitch2007} and can be used as a gauge of cluster merger activity \citep{Owers2009}. Simulations of cluster mergers have shown that cold fronts can arise in several different ways, depending on the merger conditions \citep{Poole2006,Ascasibar2006,ZuHone2010, ZuHone2016a}. However, cold fronts resulting from cluster mergers can generally be classified into two categories \citep{Owers2009, Markevitch2000}: the remnant core (e.g. the Bullet cluster; \cite{Markevitch2002}), and sloshing spirals (e.g. Abell 2052; \cite{Blanton2011} or Abell 98; \cite{Sarkar2023}). The remnant core scenario results in a contact discontinuity between the interface of the infalling subcluster's cool core and the hotter surrounding ICM. Cold fronts resulting from gas sloshing are typically seen in relaxed clusters or following a minor merger event and arise from the cool gas at the center being displaced from the central potential well, resulting in ``sloshing spirals'' \citep{Tittley2005, Markevitch2000,Roediger2011,Paterno-Mahler2013}. The observed morphology of sloshing depends on the sloshing direction. If the sloshing direction is closer to the plane of the sky, a spiral structure may be more evident in direct imaging. However, a morphology of concentric arcs around the cluster center is more likely to be observed when the sloshing direction is not in the plane of the sky \citep{Gastaldello2013}.

The study of clusters hosting sloshing spirals is important not only for informing the merger history of these clusters but because they also can provide insight into the origin of non-cool core (NCC) clusters. It is expected that cool core (CCs) systems should be host to sloshing cold fronts \cite{Markevitch2007}, however two notable exceptions to this exist. Abell 2142 \citep{Rossetti2013} and Abell 1763 \citep{Douglass2018} both lack CCs but show evidence of off-axis merger activity through the presence of sloshing cold fronts. The recent simulations of \cite{Valdarnini2021} also support this link between merger history and CC vs NCC systems. This indicates off-axis mergers may play a more meaningful role than previously assumed in the formation of NCC systems.

In addition to surface brightness edges in the hot ICM, cluster mergers can sometimes distort radio sources into a bent-tail morphology. The radio emission from these sources comes from outflows driven by the central active galactic nuclei (AGN) of the host galaxy. These outflows result in radio-emitting jets and lobes that are oriented perpendicular to the accretion disk. The tails are bent via ram pressure resulting from relative motion between the host galaxy and the ICM. One type of bent-tail radio source is the narrow-angle tail (NAT), so-called because the radio lobes are typically bent back at large angles (relative to the outflowing jet axis). NATs are generally thought to be formed via ram-pressure resulting from the high-speed infall or orbital motions of the host galaxy relative to the ICM. However, \cite{Bliton1998} suggest that, in many cases, the orbital velocity of the host galaxy is simply not high enough to produce the high degree of bending seen in NATs. They present an alternative scenario of NAT formation due to the host cluster experiencing a merger event. In this scenario, the cluster merger activity induces bulk motion of the ICM, which has been shown to produce the ram-pressure needed to bend the radio jets to the high angles of NATs. This has since been supported by simulations which show that high-velocity bulk flow of the ICM can produce bending of radio tails \citep{Roettiger1993,Mendygral2012}.

In this paper, we present \chandra X-ray observations of the galaxy cluster Abell 119 (z=0.044; A119 hereafter). A119 lacks a cool core \citep{Lagana2019} and previous studies have reported global X-ray temperatures ranging from 5.1 to 5.9 keV \citep{Edge1990,David1993,Ebeling1996, Markevitch1998,Elkholy2015}. The brightest cluster galaxy, UGC 579, is classified as a cD galaxy \citep{Postman1995, Saglia1997}. A119 hosts two NAT sources located E (0053-015), which we will refer to as the ``eastern'' NAT, and SW (0053-016), which we will refer to as the ``western'' NAT, of the central cD galaxy \citep{Feretti1999}. There is a third radio source, 3C 29, located $\sim 22$\arcmin\ from the cluster center, but lies outside of our Chandra coverage and is therefore excluded from the analysis in this paper. Previous studies have suggested that A119 is a dynamically complex cluster, showing evidence of multiple substructures \citep{Fabricant1993,Ramella2007, Tian2012,Lee2016}. This paper aims to use the presented \chandra X-ray observations, in conjunction with results from previous optical and radio studies, to form a picture of the merger history of A119.

Throughout this paper, we adopt a standard $\Lambda$CDM cosmology with $H_0 = 70$ km s$^{-1}$ Mpc$^{-1}$, $\Omega_M = 0.3$, and $\Omega_{\Lambda} = 0.7$. At $z=0.044$, the luminosity distance is $D_L = 194.8$ Mpc, the angular size distance is $D_A = 178.7$ Mpc, and $1\arcsec = 0.866$ kpc. Uncertainties reported here are 1$\sigma$ confidence intervals unless otherwise noted. 

\section{\chandra Observations \& Data Reduction}

We used two \chandra pointings taken 2003 September 4 (ObsID 4180; 11.93 ks) and 2007 August 20 (ObsID 7918; 45.04 ks). Both observations were taken using the ACIS-I configuration in very faint (VFAINT) mode. The cluster emission fills the majority of the four ACIS-I CCDs, so both observations also include data from the S2 CCD for use in background flare removal. 

The data were processed using CIAO (v4.12) and CALDB  (v4.9.0) provided by the \chandra X-ray Center (CXC). Level-2 events files were produced using \verb|chandra_repro| with \verb|check_vf_pha=yes|. Background flares were removed by extracting light curves using \verb|dmextract|, and detecting and removing flares using \verb|deflare|. After filtering for flares, the total exposure time of the two observations was 46.9 ks. Blank-sky background files were selected from CALDB, using \verb|acis_background_lookup|, and reprojected to match the observations. Background images were exposure corrected and normalized such that the hard band count rate in the 10-12 keV band matched that of the observations. Aspect histograms, instrument maps, and exposure maps for each observation were created following the step-by-step guide outlined in the CIAO thread \emph{Multiple Chip ACIS Exposure Map}\footnote{\href {https://cxc.harvard.edu/ciao/threads/expmap_acis_multi/}{https://cxc.harvard.edu/ciao/threads/expmap\_acis\_multi/}}. Exposure maps were then filtered to exclude regions with $<10\%$ the maximum exposure. Both ObsIDs were stacked and reprojected to match the coordinates of ObsID 7918 using \verb|reproject_events| to create a final combined image with a cleaned exposure time of 46.9 ks.

\begin{figure*}
\plottwo{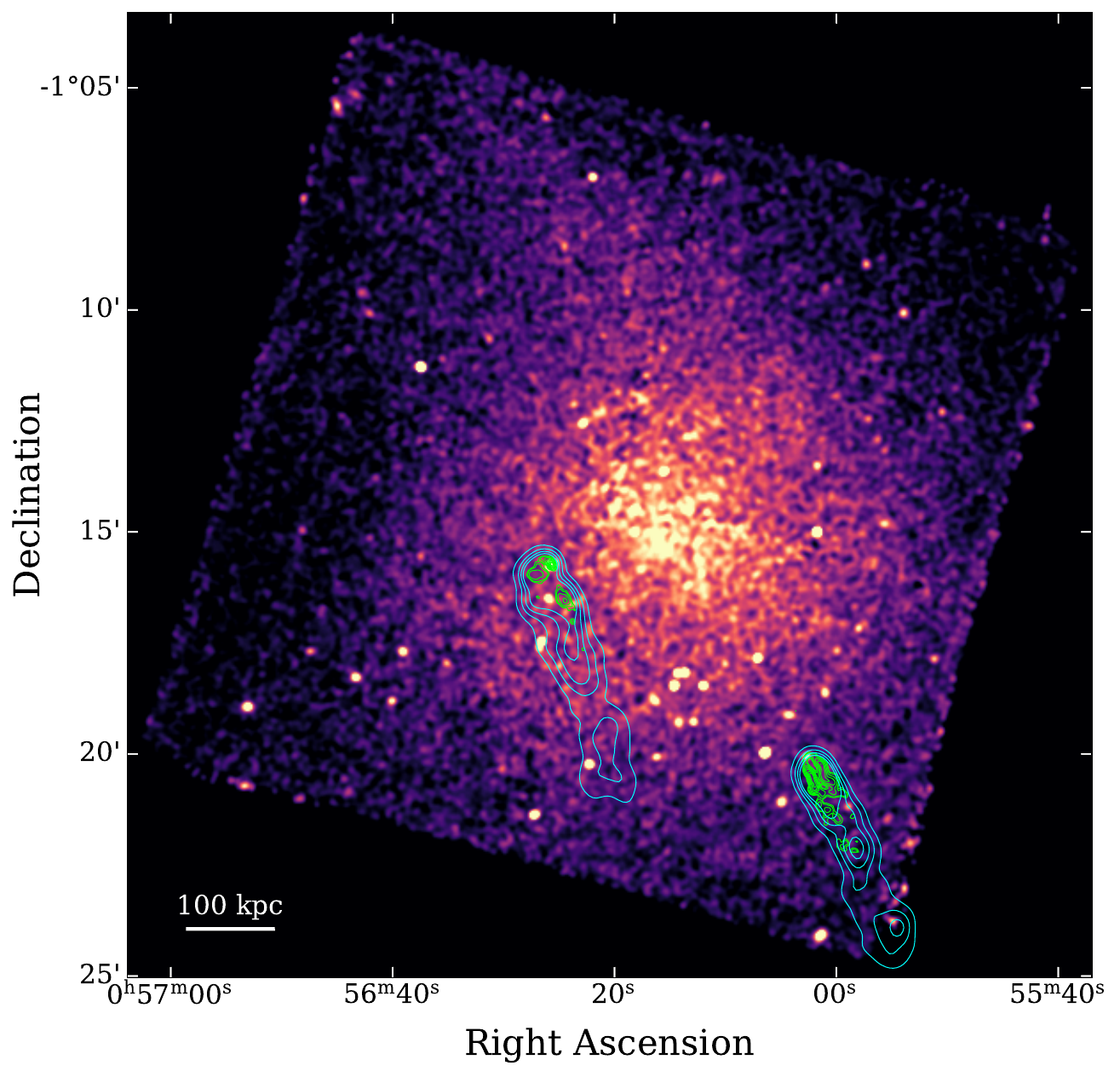}{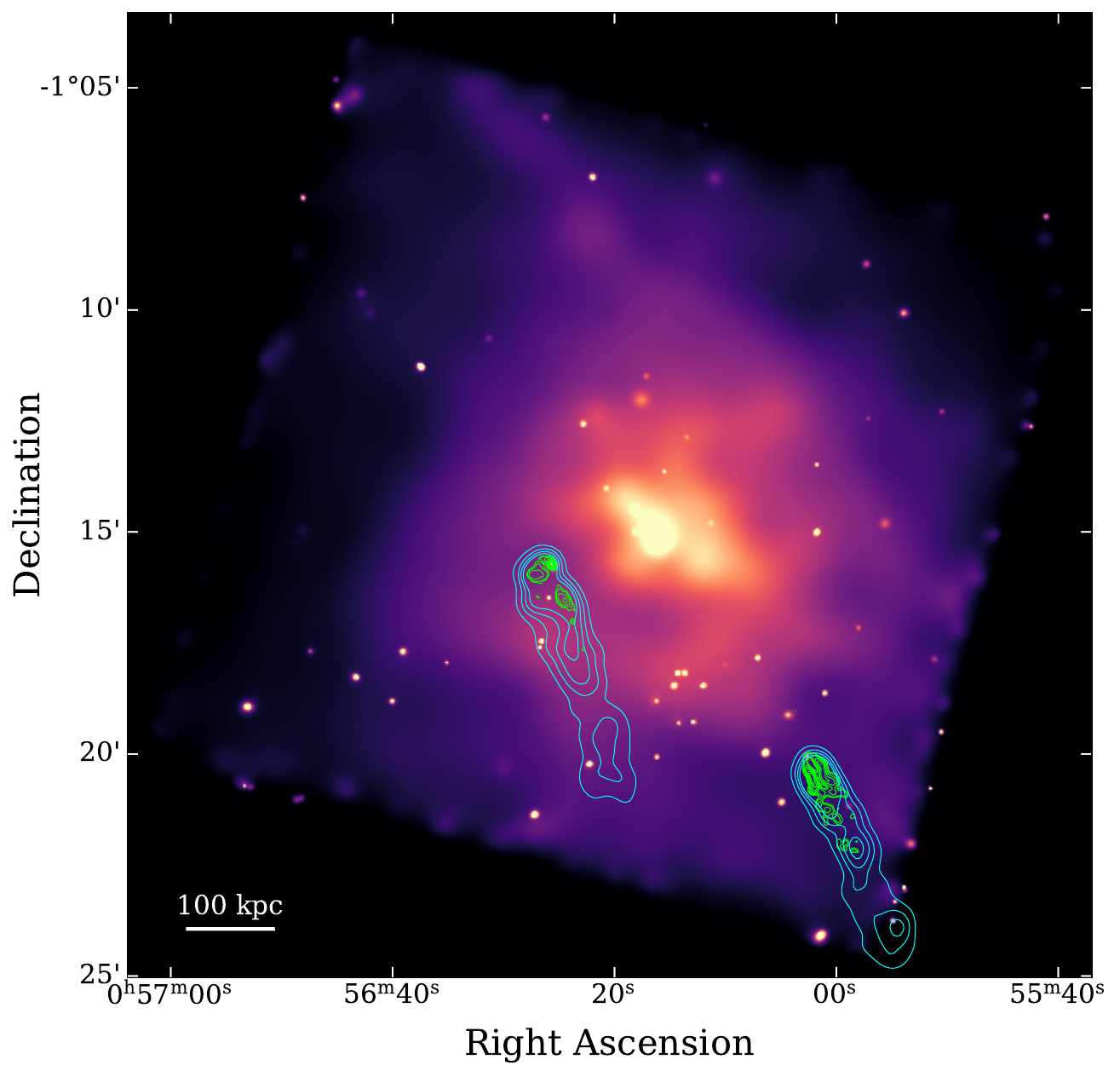}
\caption{(\emph{Left}) Gaussian smoothed (using a 7.38\arcsec\ radius Gaussian) and (\emph{Right}) adaptively smoothed background and exposure corrected composite X-ray images in the 0.7-7 keV energy range with 1.4 GHz VLA FIRST (green) and 150 MHz TGSS (cyan) radio contours.  \label{fig:xray}}
\end{figure*}

\begin{figure*}
\begin{center}$
\begin{array}{cc}
\includegraphics[width = 0.3 \textwidth]{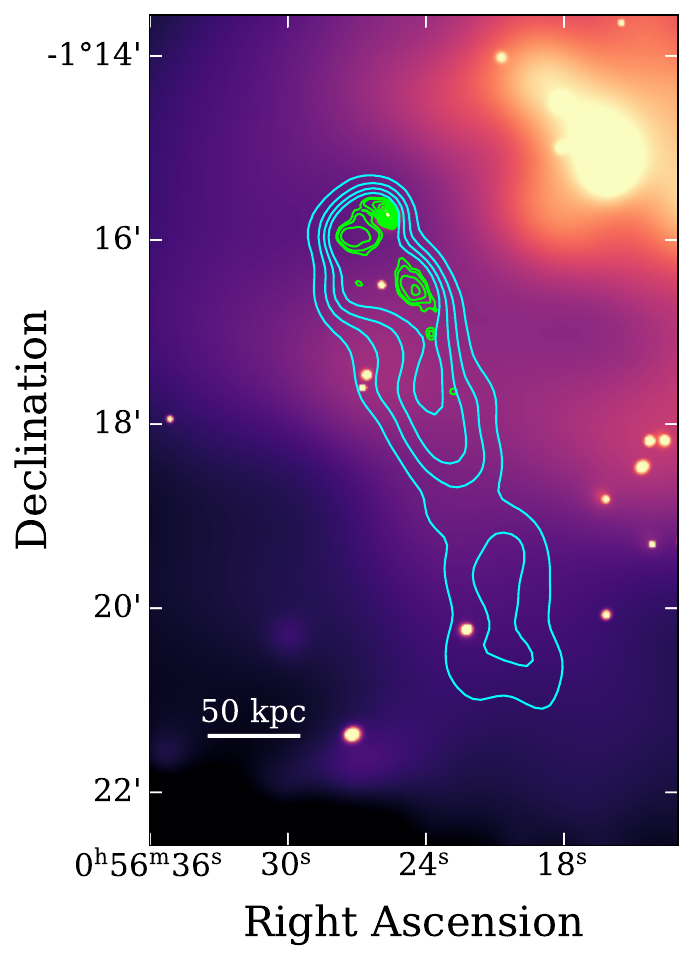} & 
\includegraphics[width = 0.35 \textwidth]{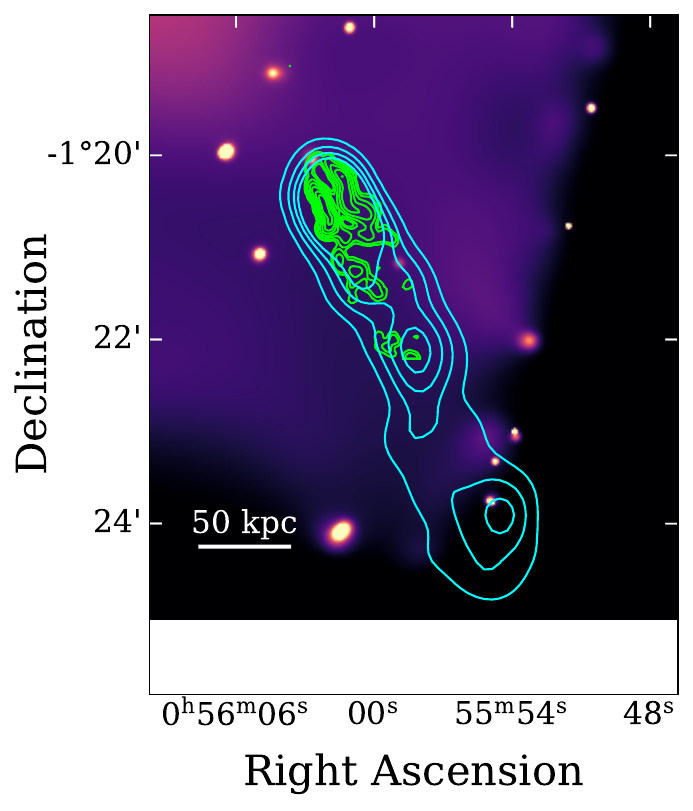}  
\end{array}$
\caption{Zoom in of the eastern (\emph{Left}) and western (\emph{Right}) NAT sources on the adaptively smoothed X-ray image. Overlaid are the 1.4 GHz VLA FIRST (green) and 150 MHz TGSS (cyan) radio contours.}
\end{center}
\label{fig:nats}
\end{figure*}

\section{Imaging Analysis \label{sec:imaging}}

The left panel of Fig.\ \ref{fig:xray} shows the background-subtracted and exposure-corrected composite \chandra image in the 0.7-7 keV band. The image was smoothed using a 7.38\arcsec\ radius Gaussian. In addition to Gaussian smoothing, we use the CIAO routine \verb|csmooth|, with a minimum signal-to-noise ratio of 4, to adaptively smooth the X-ray image. The right panel of Fig.\ \ref{fig:xray} shows the adaptively smoothed, background and exposure corrected composite X-ray image of A119 in the 0.7-7 keV band. The overall X-ray emission is fairly asymmetric with an elongation to the NE resulting in a ``teardrop'' shape. The adaptively smoothed X-ray image shows the possible presence of clumpy substructure within the ICM. 

Overlaid on the two X-ray images are radio contours of the two NAT sources. The green contours are taken from the Very Large Array (VLA) Faint Images of the Radio Sky at Twenty-cm (FIRST) survey at 1.4 GHz with a beam size of $6.4\arcsec\times 5.4\arcsec$ \citep{Becker1995}. The cyan contours are taken from the 150 MHz all-sky survey observed with Giant Metrewave Radio Telescope (GMRT) as part of the TIFR GMRT Sky Survey (TGSS) project with a $26.7\arcsec\times 24.9\arcsec$ beam size \citep{Intema2017}.

\begin{figure*}
\begin{center}$
\begin{array}{cc}
\includegraphics[width = 0.42 \textwidth]{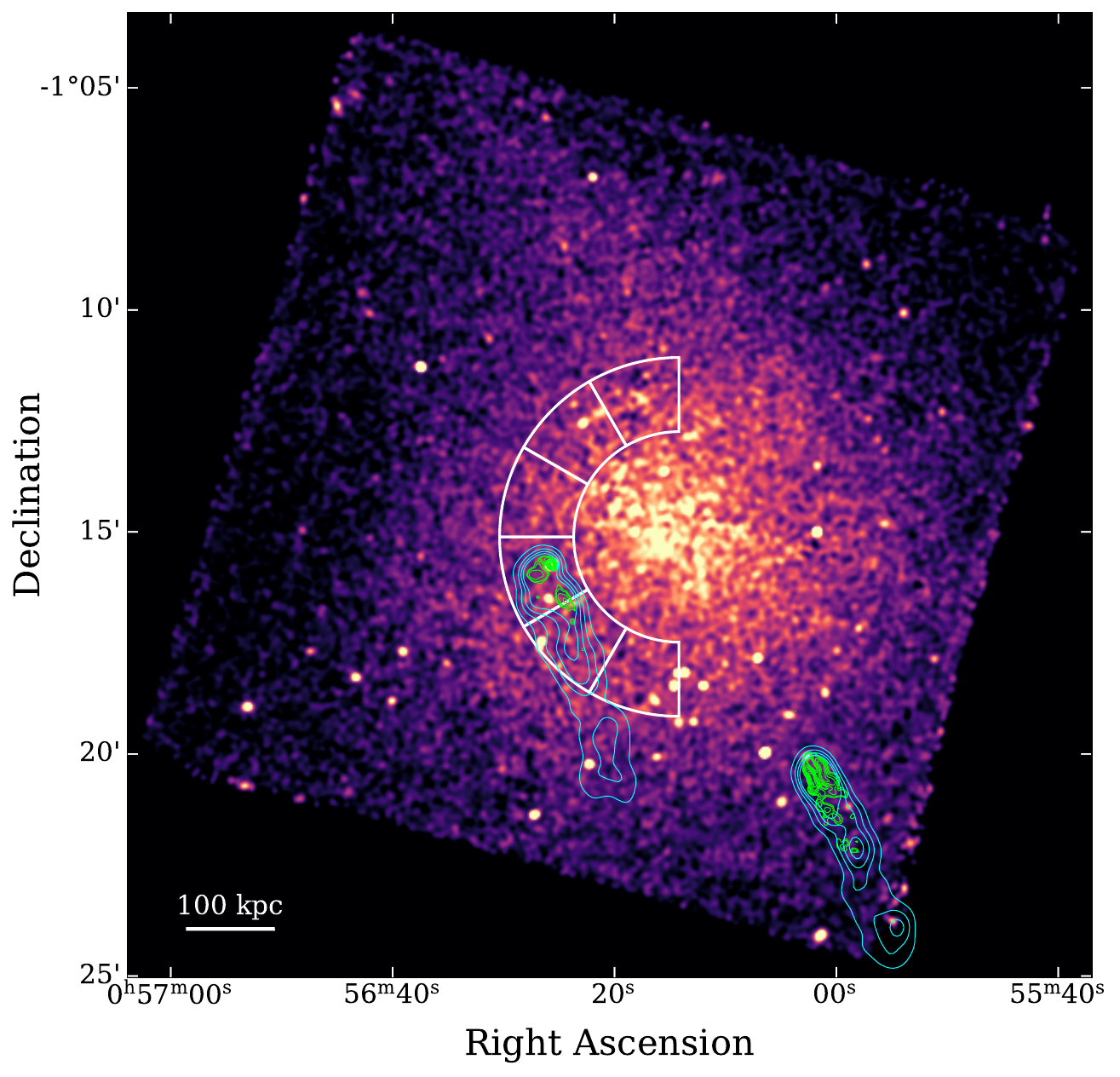} & 
\includegraphics[width = 0.45 \textwidth]{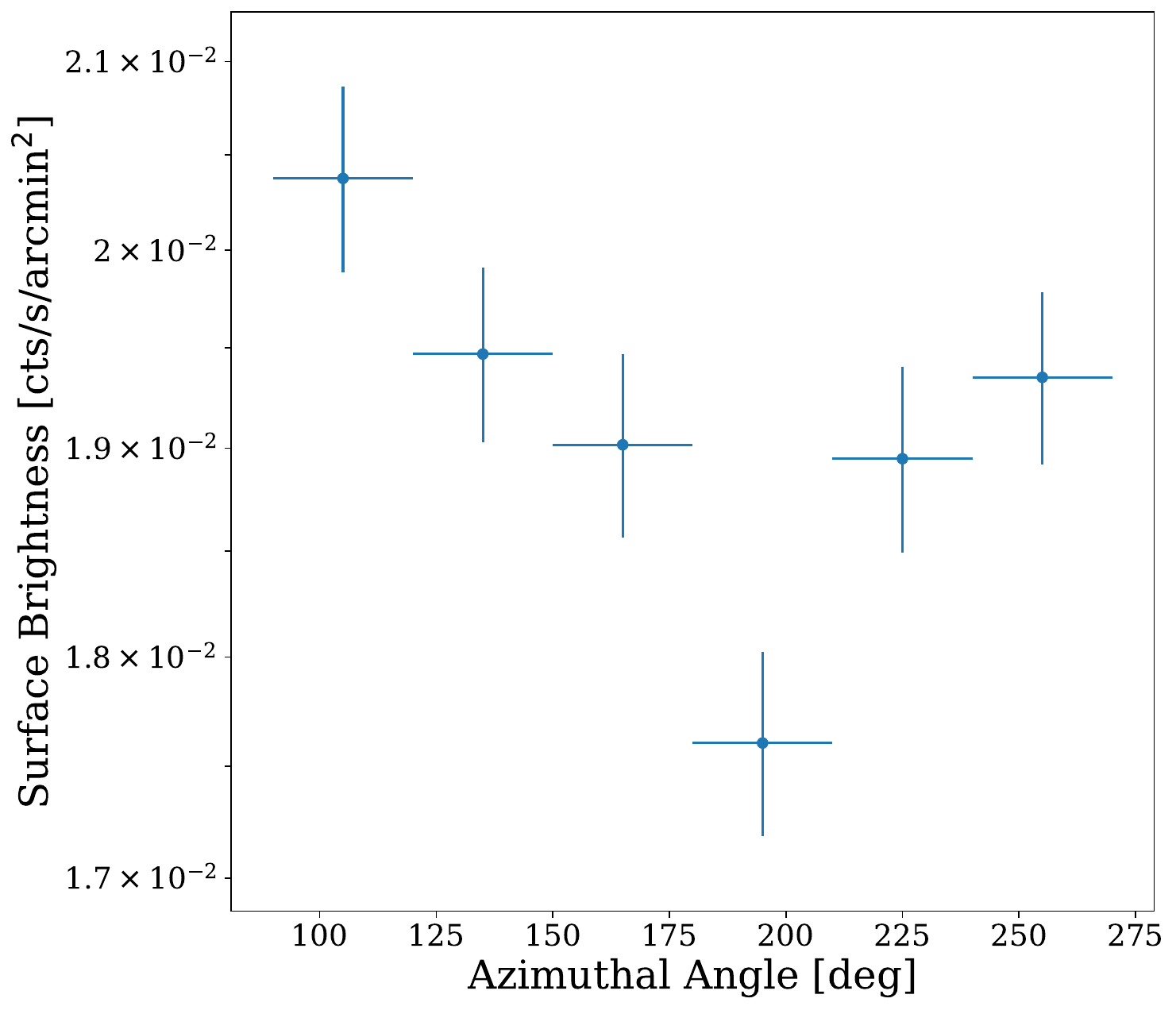} \\ 
\end{array}$
\caption{(\emph{Left}) X-ray image showing the region used to plot the azimuthal surface brightness profile (\emph{right}) across the eastern NAT source to highlight the decreased X-ray emission surrounding the NAT discussed in \S\ref{sec:imaging}.}
\end{center}
\label{fig:azprof}
\end{figure*}

Fig.\ \ref{fig:nats} shows zoomed in views of the adaptively smoothed X-ray image in the region of the two NAT sources with the 150 MHz and 1.5 GHz radio contours overlaid, showing the jet/lobe structure in more detail. The tails of the two NAT sources are also orientated along the NE-SW direction, following the X-ray emission. This is notable because while the tails are oriented parallel, the jets actually leave their host galaxies in very different directions. In the eastern NAT, the jets are leaving the host galaxy along the NE-SW plane, while in the western NAT the jets are leaving in more of the SE-NW direction. The 150 MHz contours of the eastern NAT show a break in the radio emission that appears to bisect an X-ray brightness edge to the south. In addition, the eastern NAT could be displacing the surrounding X-ray emitting gas. There is a prominent cavity of decreased surface brightness around the NAT that can be seen in the adaptively smoothed X-ray image in Fig.\ \ref{fig:xray} and the zoomed in view in Fig.\ \ref{fig:nats}. The region of decreased X-ray surface brightness is localized to the region of higher frequency radio emission, shown by the green contours in Fig.\ \ref{fig:nats}. To asses the significance of the cavity, we plot the surface brightness in a semi-circular region over the NAT source (see Fig.\ \ref{fig:azprof}). The bubble of decreased emission is found to be significant at the 3.3$\sigma$ level, relative to the average background of the two adjacent bins. The interior edge of the X-ray deficient cavity is also observed as a surface brightness edge at a radius of $\sim 135$\arcsec\ in the radial profiles across the NAT source, which are discussed later in \S \ref{sec:vel} and shown in Fig.\ \ref{fig:nat15}.

\begin{figure}
    \plotone{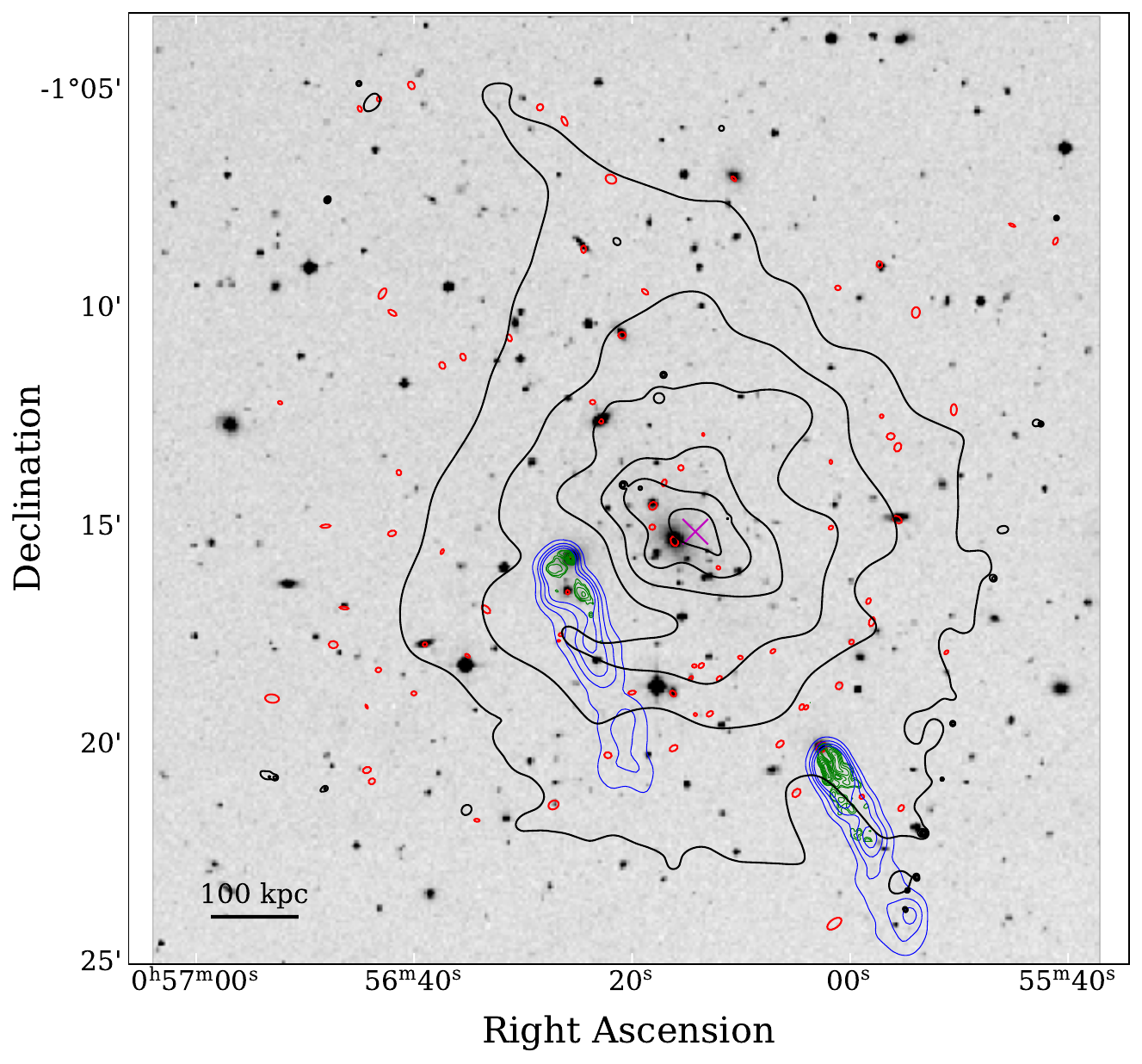}
    \caption{Optical image from the Digitized Sky Survey (DSS) with the 1.4 GHz (green) and 150 MHz (blue) radio contours as well as black contours showing the X-ray emission from the adaptively smoothed image. Red ellipses mark point sources detected by \texttt{wavdetect}. The magenta `x' marks the peak of the X-ray emission.}
    \label{fig:opt}
\end{figure}

Fig.\ \ref{fig:opt} shows an optical image from the Digitized Sky Survey (DSS) with contours of X-ray emission (black), taken from the adaptively smoothed X-ray image (right panel of Fig.\ref{fig:xray}), and radio (green and blue) emission of the two NAT sources are overlaid. The magenta `x' in the optical image marks the peak of the X-ray emission and shows that the central cD galaxy is slightly offset to the east of the peak. Point sources are identified using the CIAO routine \verb|wavdetect|, using wavelet scales of 4, 6, 8, and 16 pixels (1 pix = 0.492\arcsec). The detected point sources are indicated in Fig.\ \ref{fig:opt} with red ellipses. After removing the detected point sources, we use the CIAO routine \verb|dmfilth| to replace the removed point sources pixels with interpolated pixel values from the surrounding area, creating an image of the diffuse emission which is used in some of the following analysis. 

\subsection{Highlighting X-ray Substructure}

To enhance features seen in the X-ray image, we employed two techniques: beta-model subtraction and Gaussian Gradient Magnitude (GGM) filtering.  For both methods, we use the unsmoothed, merged, background and exposure corrected 0.7-7 keV image of diffuse emission. These techniques help to highlight faint structure within the ICM and are briefly described below.

\subsubsection{Beta-model Residual Image}

We fit an elliptical 2D $\beta$-model using the \verb|beta2d| model in \verb|Sherpa| to a Gaussian smoothed ($\sigma=2\arcsec$) background and exposure corrected image in the 0.7-7 keV range. The peak of the X-ray emission, as determined by the 2D $\beta$-model fit, is found to be located roughly 41 kpc NNW of the central cD galaxy (see Fig.\ \ref{fig:opt}). 

\begin{figure}
\plotone{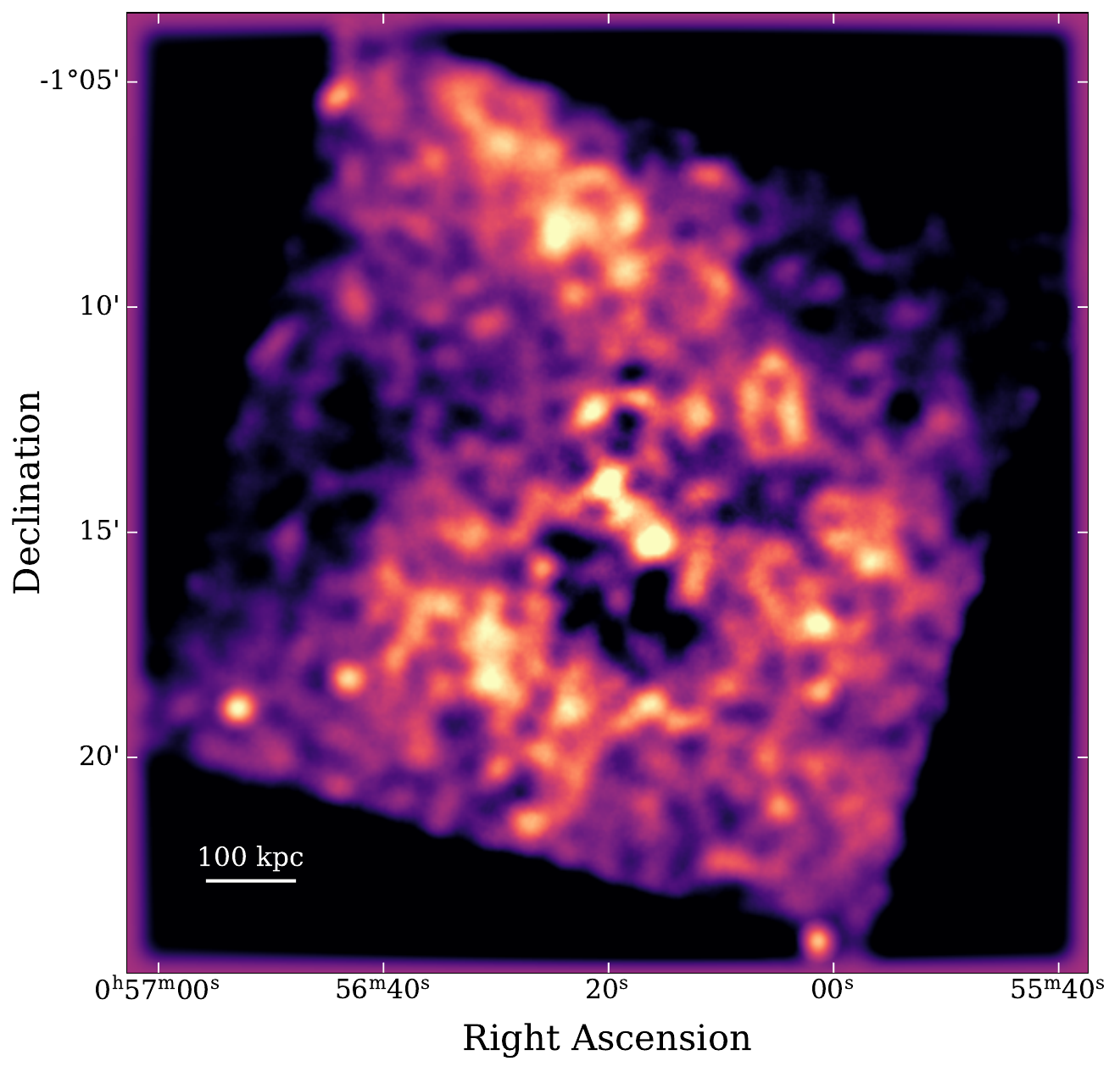}
    \caption{Excess emission in the 0.7-7 keV energy range after subtraction of a 2D $\beta$-model. The residual image was smoothed using a $\sigma = 15\arcsec$ Gaussian.}
    \label{fig:beta2d}
\end{figure}

The fitted 2D $\beta$-model is subtracted from the unsmoothed X-ray image and the resulting residual image, smoothed using a $\sigma = 15\arcsec$ Gaussian, is shown in Fig.\ \ref{fig:beta2d}. The smoothed residual image shows the presence of brightness edges seen in the Gaussian smoothed and adaptively smoothed X-ray images (Fig.\ \ref{fig:xray}). The model subtracted residual image shows a similar tear-drop like shape seen in the X-ray image, with excess emission extending over 580\arcsec to the NNE from the cluster center. Additionally, excess emission towards the south suggest the presence of an X-ray brightness edge along the southern edge of the cluster. These features are studied in more detail in \S \ref{sec:radprof}.

\subsubsection{Gaussian Gradient Magnitude (GGM) Filter\label{sec:ggm}}

\begin{figure}
\plotone{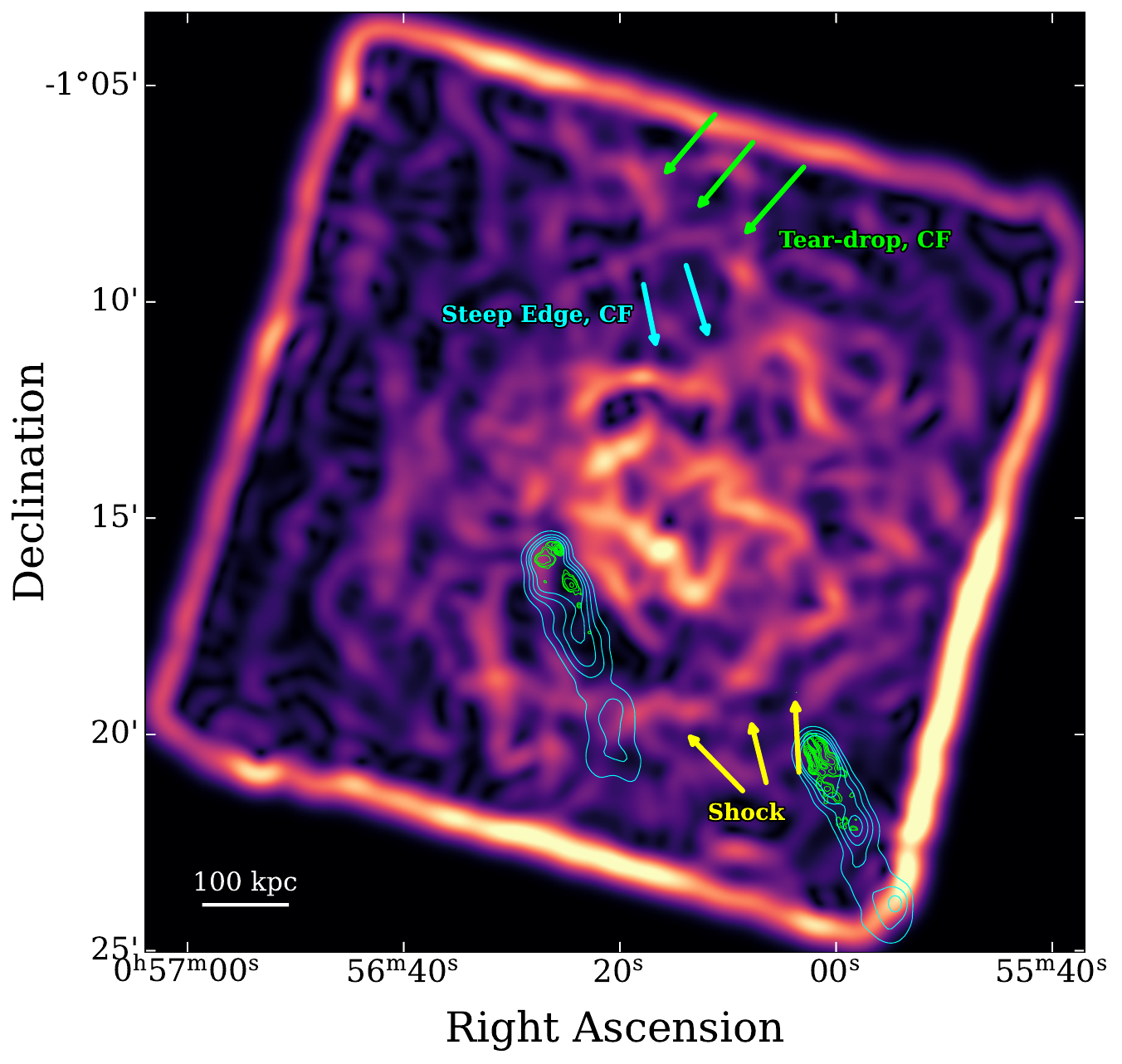}
\caption{Gaussian gradient magnitude filtered (0.7-7 keV) image with $\sigma = 40$ pixels (1 pixel $=$ 0.492\arcsec). The brightest regions correspond to the steepest changes in X-ray surface brightness. Overlaid are the 1.4 GHz (green) and 150 MHz (cyan) radio contours. Some notable features are highlighted.}
\label{fig:ggm}
\end{figure}

A Gaussian Gradient Magnitude (GGM) filter was applied using the \verb|gaussian_gradient_magnitude| within \verb|SCIPY|\footnote{http://scipy.org/}. GGM filtering calculates gradients in the 2D data, assuming Gaussian derivatives, thereby enhancing surface brightness edges \citep{Sanders2016}. The scaling is determined through the choice of $\sigma$ (the width of the Gaussian assumed), with smaller values typically chosen to highlight small scale structure while larger values of $\sigma$ can be used to highlight large scale structure. Figure \ref{fig:ggm} shows the GGM filtered image with $\sigma = 40$ pixels (1 pixel = 0.492\arcsec). This scaling was chosen to highlight large-scale structures, particularly in the cluster outer regions where brightness edges are seen in the X-ray image.

While we perform a more detailed analysis of the significance of the brightness edges in \S \ref{sec:radprof}, here we highlight some of the features seen in the GGM filtered image. Several notable features are indicated in Fig.\ \ref{fig:ggm}. In the far north, there is an edge along the western part of the tear-drop shape (green arrows). North of the cluster center there is an edge that has a concave shape (cyan arrows), similar to the Kelvin-Helmholtz instabilities that have been seen in earlier observations (e.g., \cite{Walker2017}) and in simulations of sloshing spirals \citep{Roediger2013}. To the south, there is an edge (yellow arrows) which could be bisecting the southern end of the eastern NAT's tail. A more detailed analysis of the surface brightness edges identified in A119 is presented in \S\ref{sec:radprof}.

\section{Spectral Analysis \label{sec:spec}}

In order to determine the thermodynamic properties of the ICM in A119, we extract the X-ray spectra from our \chandra observations. Unless otherwise noted, all spectra were extracted separately for the two observations using \verb|specextract| with point sources excluded. Background spectra were taken from the hard band normalized, reprojected blank-sky background observations. Spectra were binned such that each energy bin contained a minimum of 40 background subtracted counts. Extracted spectra were fitted simultaneously in the 0.7-7 keV range in \verb|XSPEC| \citep{Arnaud1996} using a thermal \verb|APEC| model. Galactic absorption was fixed at $N_H = 3.2\times 10^{20}$ cm$^{-2}$ \citep{Dickey1990}. We adopted the \cite{Asplund2009} solar abundance tables. The temperature, metallicity, and normalization were allowed to vary.

\subsection{Total Spectrum \label{sec:totalspec}}

To determine global ICM properties we extracted the spectrum from a 400\arcsec\ circular region, centered on the cluster center. This was the largest region size that could be used before reaching a chip edge. From a total of $\sim$ 87,000 background subtracted counts, the global temperature was found to be $6.0\pm 0.1$ keV with an abundance of $0.41 \pm 0.04$ Z$_{\odot}$. The fit was good with $\chi^2 =  648.6$ with 597 degrees of freedom, corresponding to a reduced $\chi^2 = 1.1$. From the global temperature value, the average sound speed of the cluster was calculated to be $c_s = \sqrt{\gamma \frac{kT}{\mu m_p}} = 1265 \pm 11$ km/s, where $\gamma = \frac{5}{3}$, $\mu = 0.6$, and $m_p$ is the mass of the proton.  

In addition to the global fit, we created profiles of the surface brightness, projected temperature, pressure and density for the overall cluster emission. Concentric circular annuli were defined extending out to a radius of 400\arcsec\ to match the region defined above. Surface brightness profiles were extracted from the merged, background subtracted, and exposure corrected X-ray image with point sources excluded. Projected temperature and abundance profiles were created by extracting and fitting the spectrum within each circular annulus, using the method described in \S\ref{sec:spec}. 

The pressure and density profiles were calculated following the methods of \cite{Kriss1983}, \cite{Wong2008} and \cite{Blanton2009}. The X-ray surface brightness profiles were deprojected to determine the X-ray emissivity and gas density, with both assumed to be constant within spherical shells. To determine the radial variations in pressure, we used the deprojected density with the fitted projected temperatures, assuming the fitted temperature from the projected spectra is the temperature at that spherical radius. 

\begin{figure}
\plotone{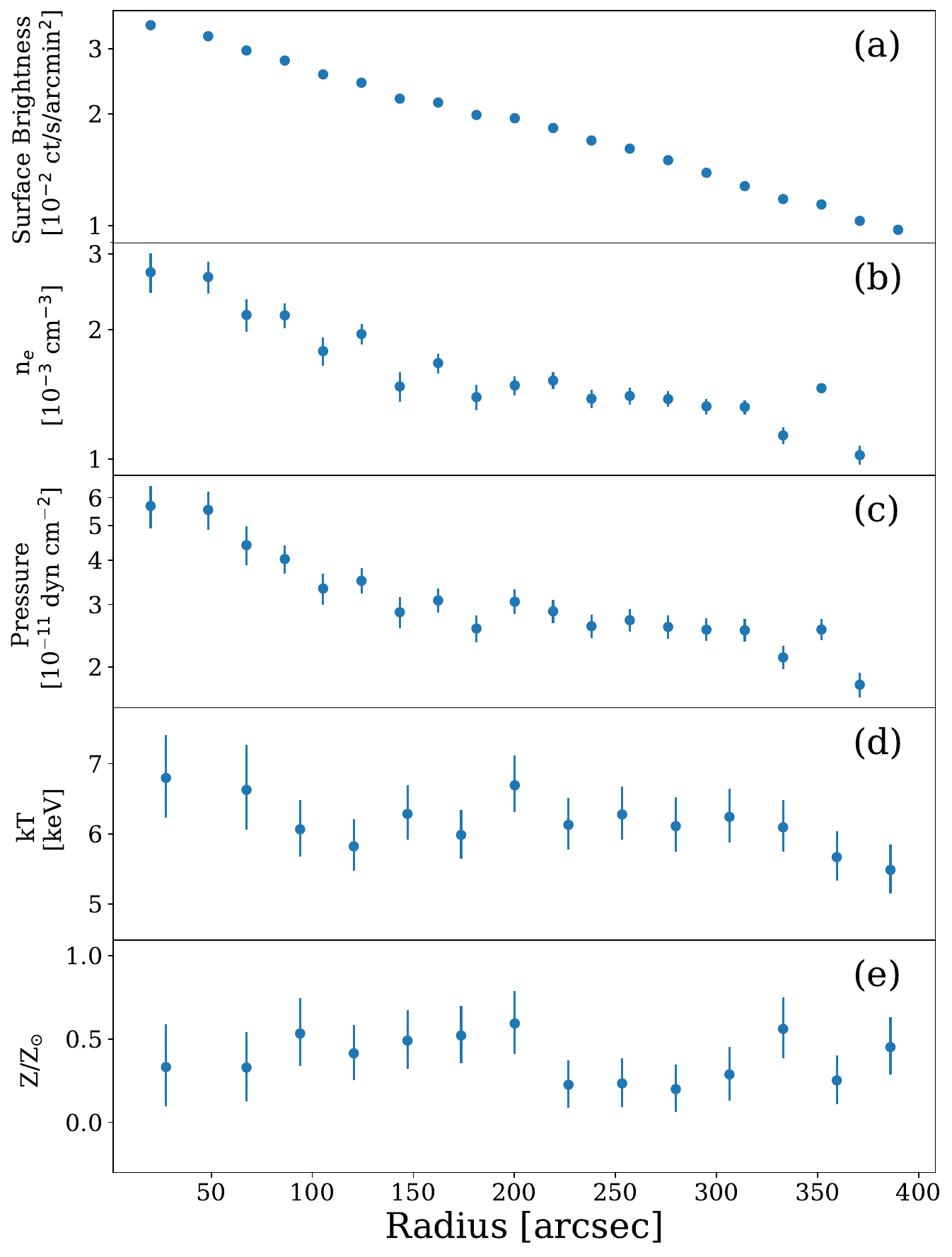}
    \caption{Radial profiles of (a) surface brightness, (b) deprojected density, (c) deprojected pressure, (d) projected temperature, and (e) projected abundance for the global cluster emission. }
    \label{fig:globalprof}
\end{figure}

Fig.\ \ref{fig:globalprof} shows the radial profiles of (a) surface brightness, (b) density, (c) pressure, (d) projected temperature, and (e) projected abundance. The temperature shows some radial variation with an overall slight decline from the center outward. The abundance profile is roughly constant with radius, with some variations, including a drop at approximately 200\arcsec. The average values of the temperature and abundance from the radial profiles is consistent with the values found above for the total spectrum fit. While the profiles of brightness, density, and pressure show some indication of a change in slope at $\sim$ 180\arcsec, which would correspond to the slight increase in temperature seen in panel (d), there is no notable presence of the edges discussed in \S \ref{sec:imaging}. Because there is no apparent indication of edge features in the global radial profiles, we define a number of subregions in which we examine the edge features seen in the X-ray images in more detail (see \S \ref{sec:radprof}).

\subsection{Temperature Map}
We created a spectral map of A119, following the methods of \cite{Randall2008, Randall2009}, to examine the temperature distribution of the cluster. For each spectral map pixel, a circular region was defined and allowed to increase in radius until a minimum of 2600 net source counts is contained. Within this circular radius, the spectrum was extracted and fitted following the method described above (see start of \S\ref{sec:spec}). 

\begin{figure}
\plotone{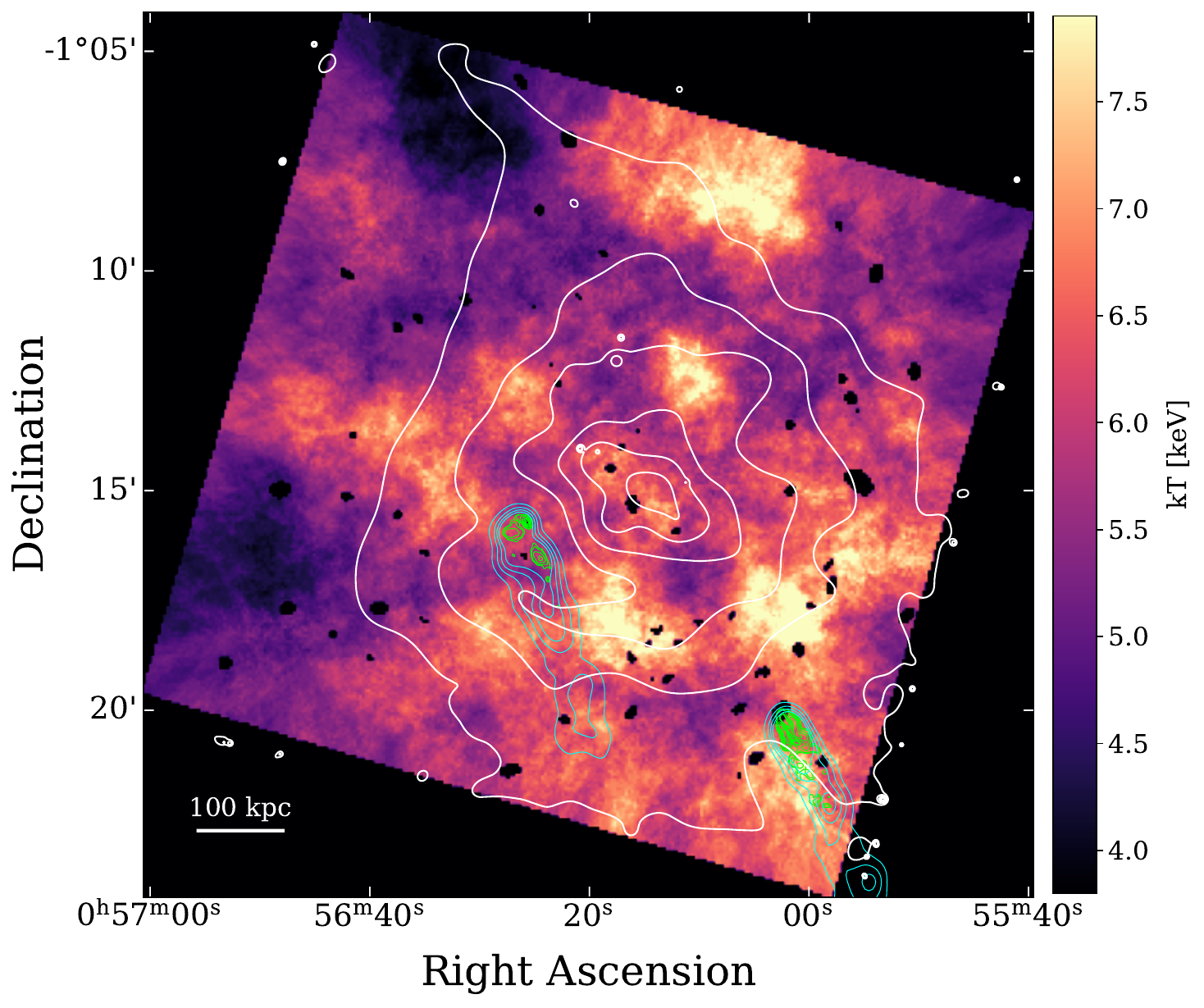}
\caption{Temperature map of A119, overlaid with 1.4 GHz (green) and 150 MHz (blue) radio contours. The white contours are from the smoothed X-ray image (see Fig.\ref{fig:xray}). }
\label{fig:tempandabund}
\end{figure}

Figure \ref{fig:tempandabund} shows the projected temperature map of A119 with 1.4 GHz FIRST and 150 MHz TGSS radio contours of the two NAT sources overlaid. The white contours are taken from the smoothed X-ray image (see Fig.\ref{fig:xray}). The temperature map is binned such that the pixel scale is 3.94"/pixel. Errors for the temperature map range from about 5\% at the cluster center to about 21\% in the outskirts. 

Cool ($\sim 6$ keV) gas surrounds the central region of the cluster and is elongated along the NE-SW direction. The cool gas extends up to the north, connecting to a large ($\sim 168$ kpc in diameter) cool ($\sim 4$ keV) spot in the NE corner. The central cool gas also appears to extend out to a cold ($\sim 4$ keV) spot in the SE corner. To the north, there is a strip of cool gas corresponding to the western edge of the tear-drop shape seen in the X-ray image. Northwest of this strip is a clump of hot ($\sim 7$ keV) gas that is 137 kpc across. Clumps of hot gas surround the cluster center at radii of $\sim 230$ kpc. The clumpy hot gas to the north of the core has temperatures of $\sim 6-7$ keV while the arm of clumpy hot gas to the south of the core has temperatures of $\sim 7-8$ keV. In order to determine the significance of the features seen in the spectral maps, follow-up detailed spectral fitting is performed on a number of regions of interest and is discussed further in \S\ref{sec:radprof}. 

\section{Radial Profiles\label{sec:radprof}}

The edges seen in the X-ray and GGM images could be indicators of cold fronts or shocks. As mentioned in the introduction, cold fronts arise at the boundaries of cool gas moving through hotter, less dense surrounding gas. Cold fronts present as drops in the X-ray surface brightness (and therefore density), with an accompanying jump in temperature. The changes in the temperature and density tend to have similar amplitudes, resulting in continuous pressure across the front. Shock fronts, on the other hand, follow the Rankine-Hugoniot jump conditions, so across the front, we would expect to see a drop in brightness, density, temperature, and pressure. 

\begin{figure*}
\gridline{\fig{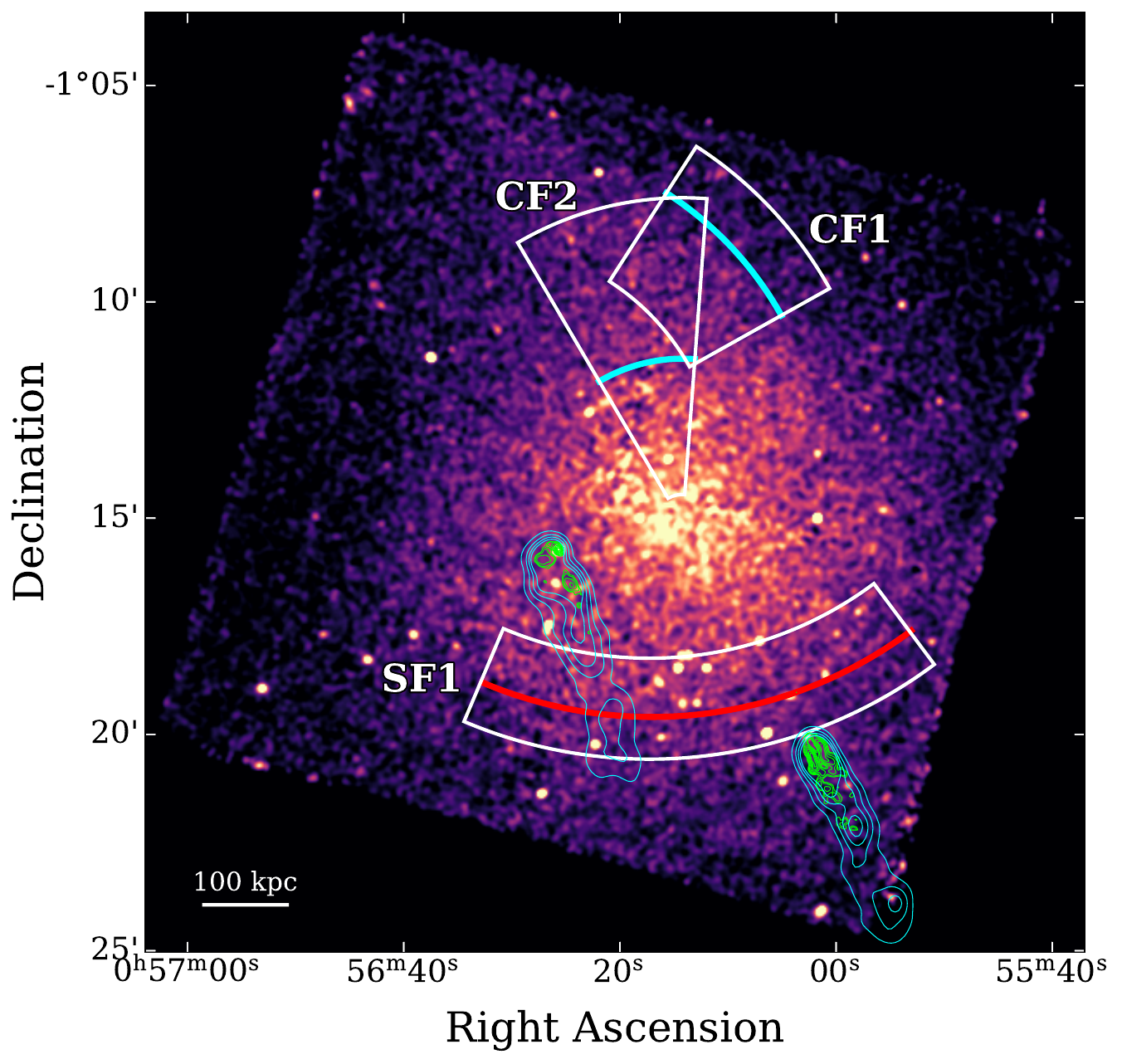}{0.5\textwidth}{}
          \fig{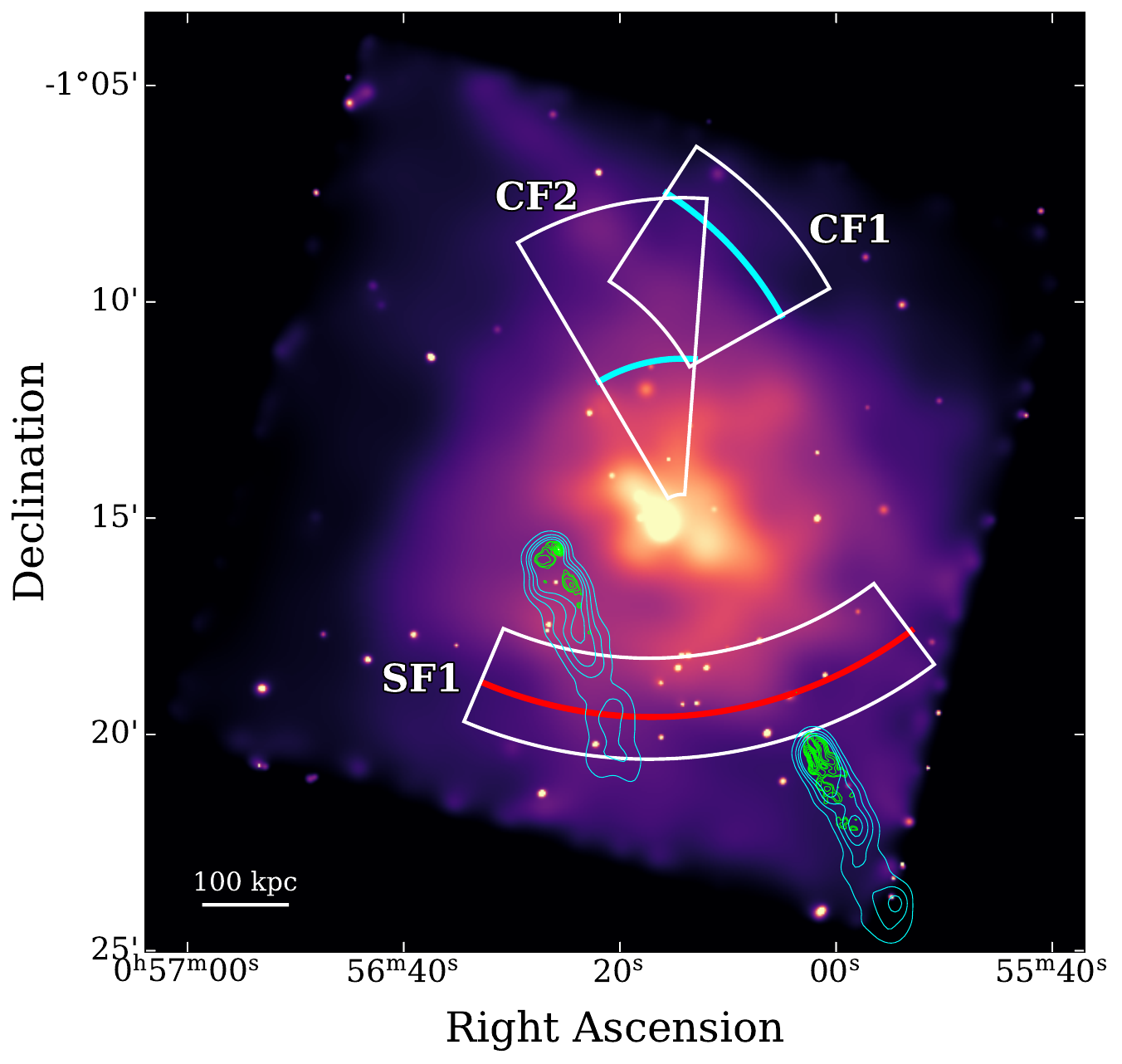}{0.5\textwidth}{}}
\gridline{\fig{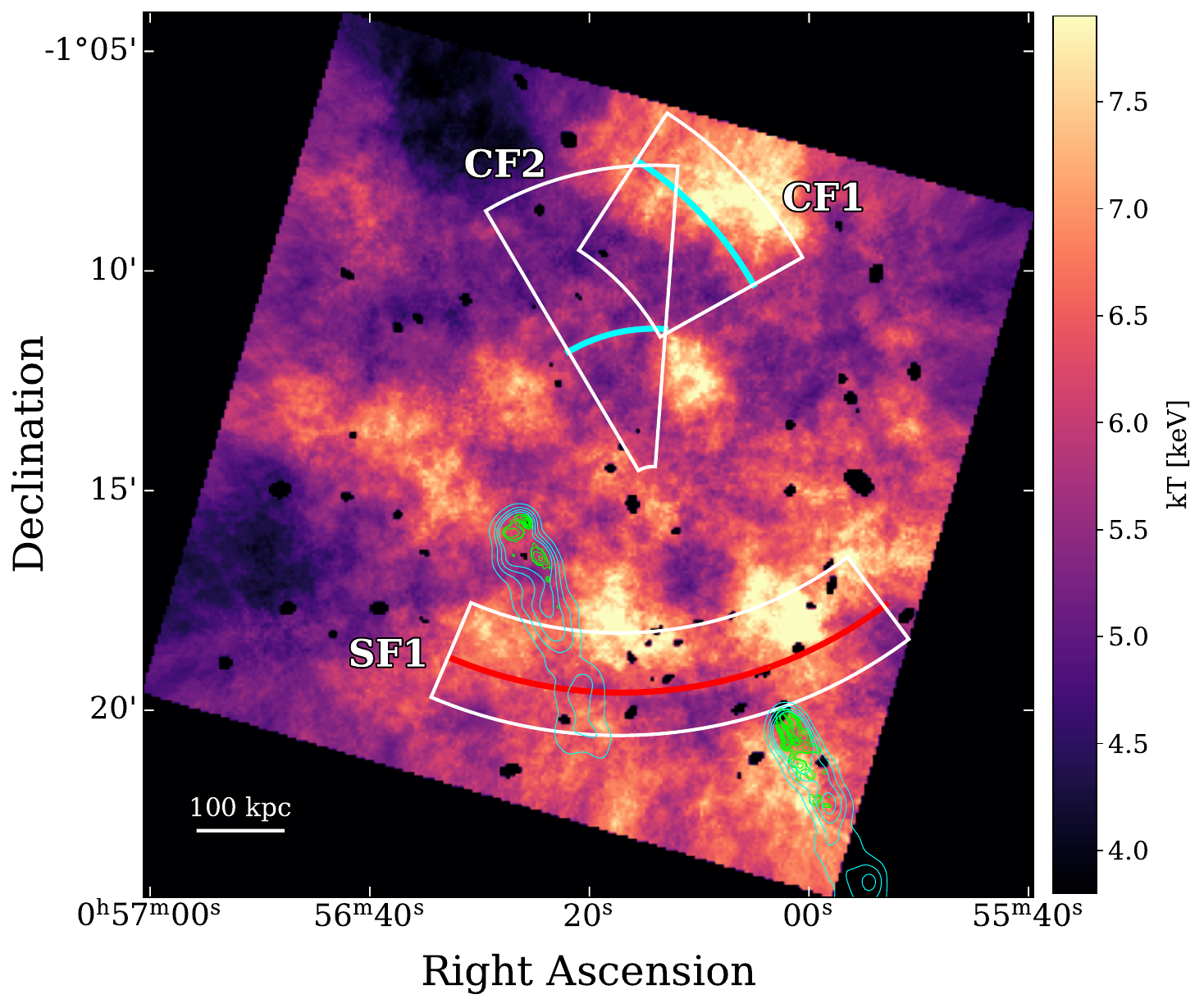}{0.53\textwidth}{}
            \fig{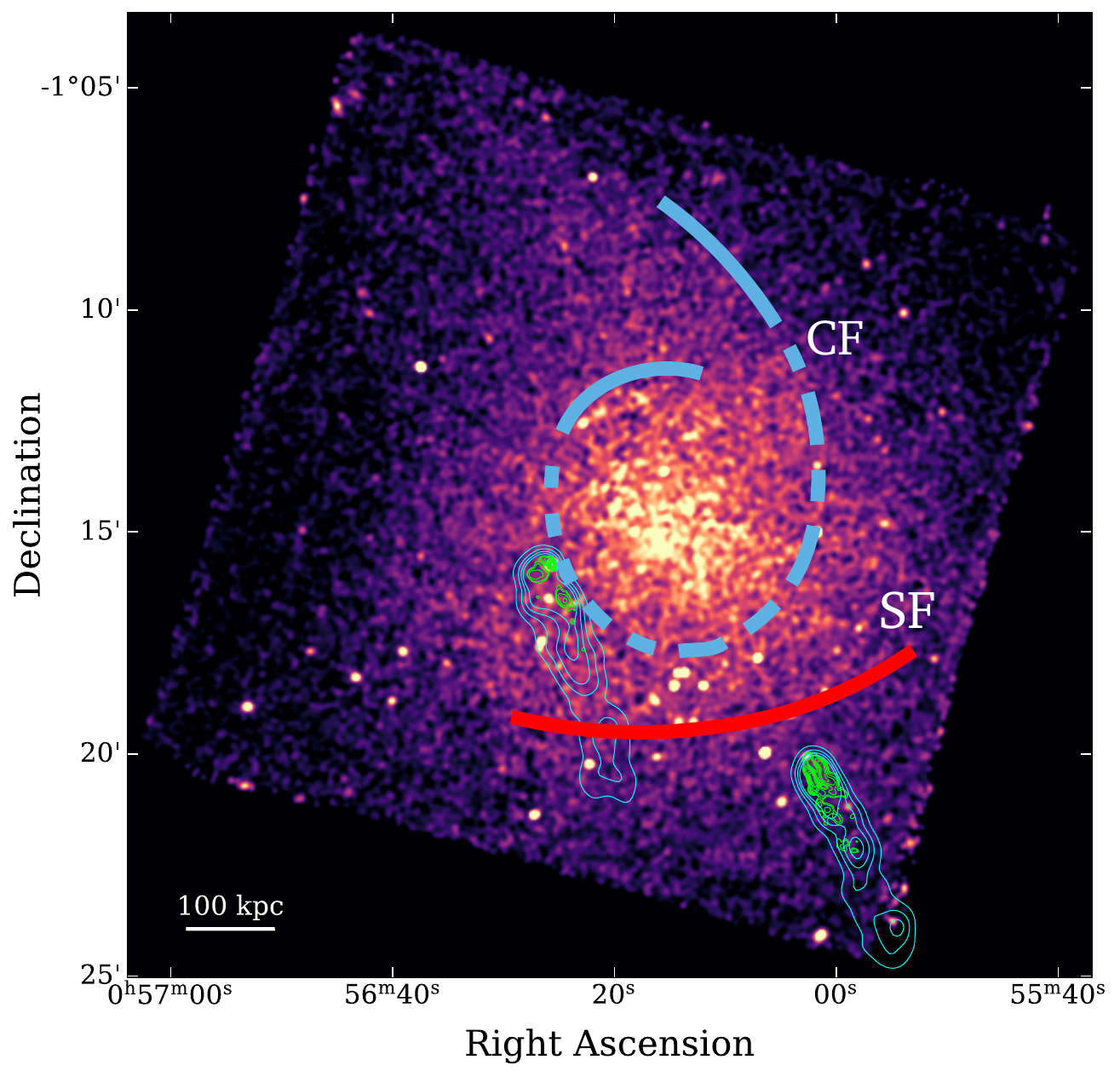}{0.47\textwidth}{}}
\caption{Gaussian smoothed (\emph{top left}; see Fig.\ \ref{fig:xray} left panel) and adaptively smoothed (\emph{top right}; see Fig.\ \ref{fig:xray} right panel) X-ray images and projected temperature map (\emph{bottom left}) showing the regions defined for extracting the radial profiles (labeled CF1-2 and SF1; see \S\ref{sec:radprof}). The solid blue (cold fronts) and red (shock) lines mark the location of the brightness edges identified from the fits to the radial profiles. In order to aid visualization, included is a cartoon schematic (\emph{bottom right}) showing the merger shock and a possible scenario of the CF regions connecting to form a sloshing spiral structure. Overlaid on all panels are the 1.4 GHz (green) and 150 MHz (blue) radio contours of the two NAT sources.\label{fig:fronts}}
\end{figure*}

To determine whether any of the X-ray structures seen in A119 are shocks or cold fronts, we defined three wedge-shaped regions for further study (shown in Fig.\ \ref{fig:fronts}). Within each wedge region, we extract surface brightness profiles in the 0.7-7 keV band from the merged, background subtracted, and exposure corrected X-ray image with point sources excluded. Projected temperature profiles were created by extracting the spectra from the same wedge regions used to get surface brightness profiles, but using fewer annuli in an attempt to minimize errors. From each annulus, the spectrum is extracted and fitted using the method described in \S\ref{sec:spec}. Pressure and density profiles were created following the deprojection method used to create the global profiles described in \S\ref{sec:totalspec}. 

We fit each surface brightness profile with a broken power-law model assuming a 3D electron density profile defined as \citep{Sarkar2022}:
\begin{equation}
n_e(r) \propto {\left\{\begin{matrix}
\left(\frac{r}{r_{\text{edge}}}\right)^{-\alpha_1} \quad \text{ if } r < r_{\text{edge}} \\ 
\frac{1}{\text{jump}}\left(\frac{r}{r_{\text{edge}}}\right)^{-\alpha_2} \quad \text{ if } r \geq r_{\text{edge}}
\end{matrix}\right.}
\end{equation}
where $n_e(r)$ is the 3D electron density at a given radius $r$, $r_{\text{edge}}$ is the distance of the putative edge from the cluster center, $jump$ is the density jump factor, and $\alpha_1$ and $\alpha_2$ are the power-law slopes inside and outside of the edge. We use this density model to estimate the emission measure profile and project it onto the sky plane, and fit the observed surface brightness profile by using the least square fitting technique. We let the parameters, $\alpha_1$, $\alpha_2$, $r_{\text{edge}}$, and the $jump$, free to vary during fitting. 

\begin{figure*}
\gridline{\fig{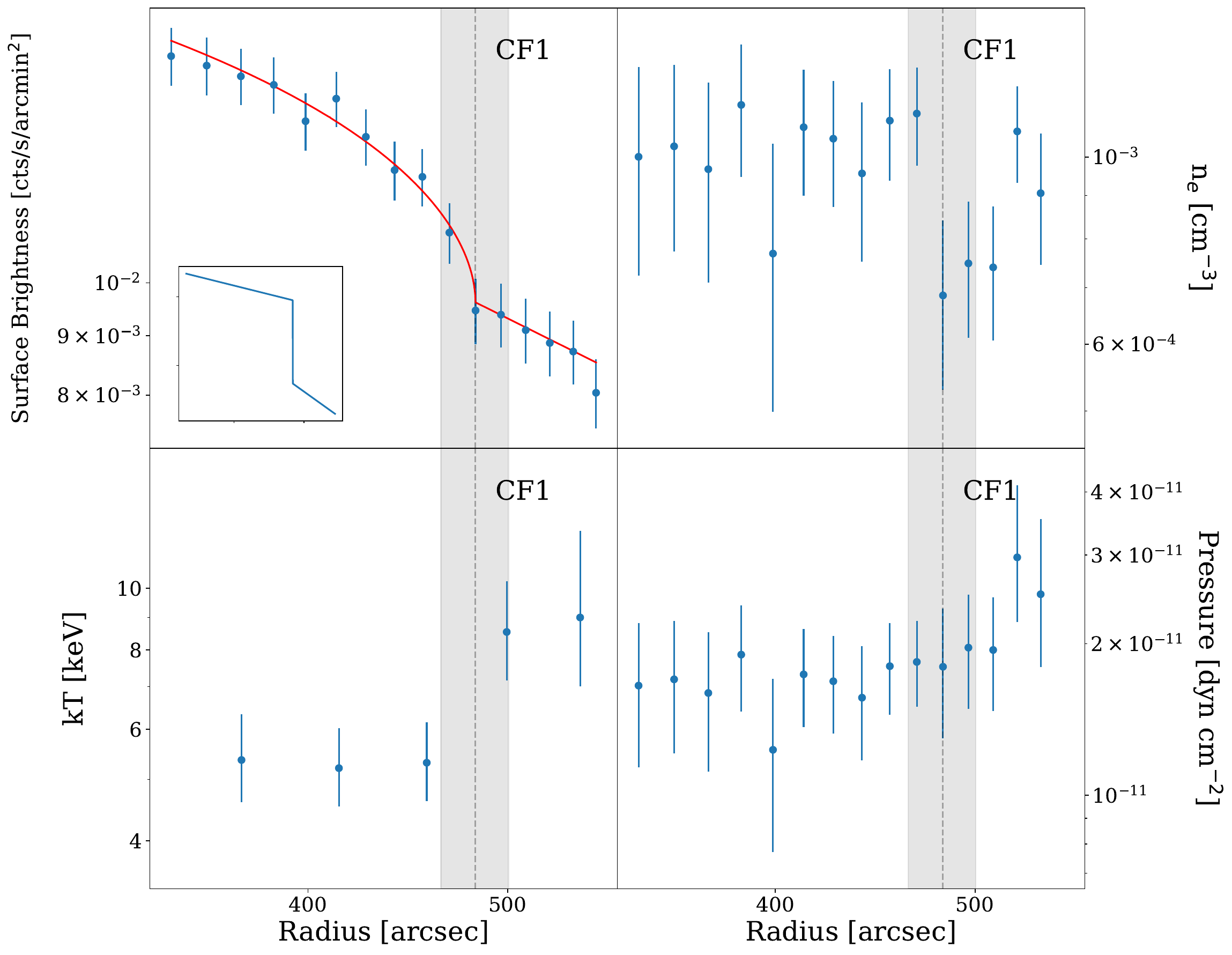}{0.5\textwidth}{}
          \fig{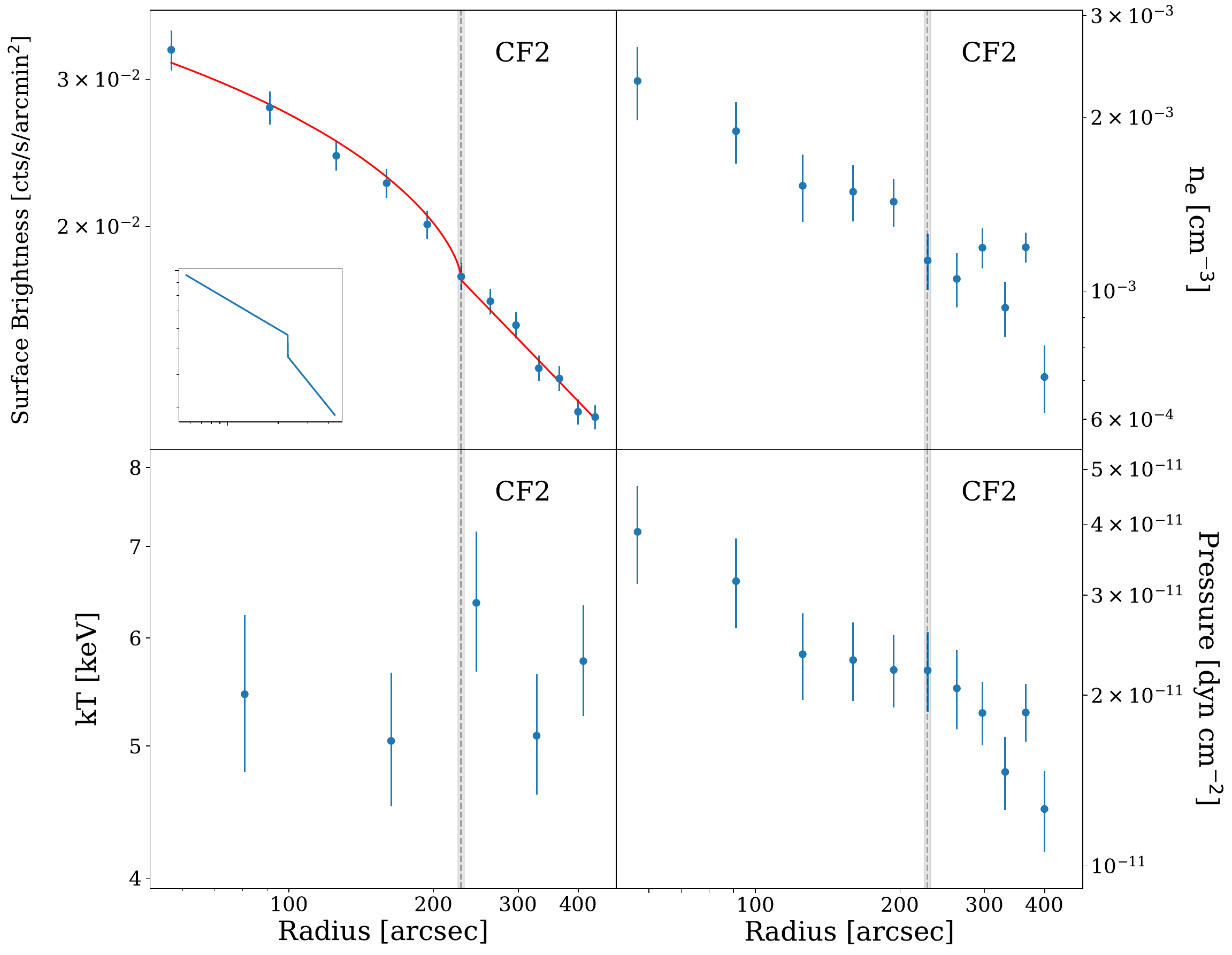}{0.49\textwidth}{}}
\gridline{\fig{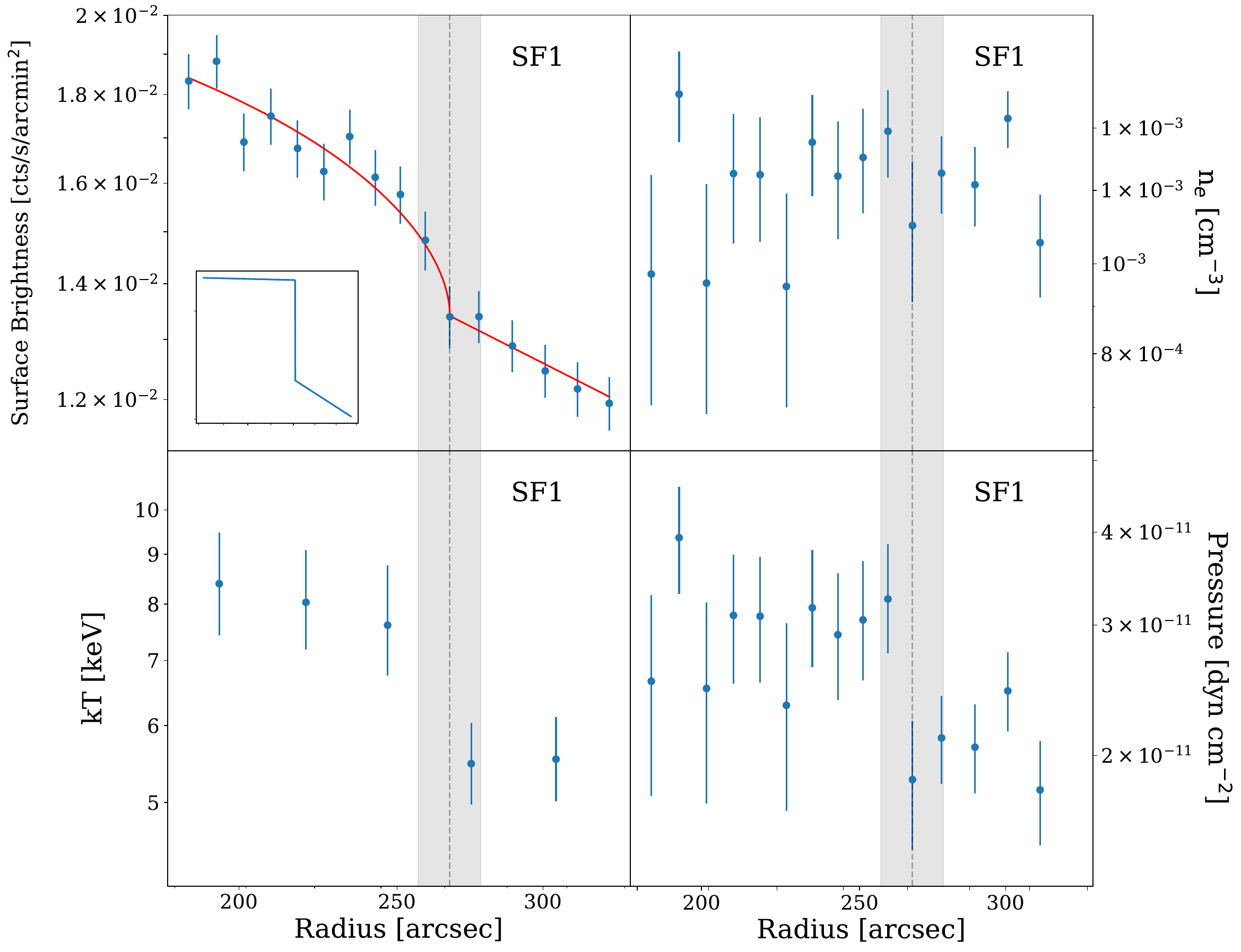}{0.5\textwidth}{}}
\caption{Radial profiles of X-ray surface brightness, projected temperature, and deprojected density and pressure for the regions discussed in \S\ref{sec:radprof} and shown in Fig.\ \ref{fig:fronts}. The surface brightness profiles include the best-fit model (red line) and the 3D density profile (inset) obtained from the fit. The dashed lines in all panels denote the best-fit edge radius and the shaded regions are the 1$\sigma$ errors on the fitted radius.}\label{fig:profs}
\end{figure*}

Figure\ \ref{fig:profs} shows the radial profiles of surface brightness, projected temperature, deprojected pressure and density, the best-fit surface brightness model, and the 3D density profiles (inset) for each region shown in Fig.\ \ref{fig:fronts}. 

\paragraph{Region CF1} The radial profiles of region CF1 show a drop in X-ray brightness and deprojected density at 470\arcsec. Across the edge, the temperature increases from \fe{5.3}{0.8}{0.7} to \fe{8.5}{1.7}{1.4} while the pressure remains constant. The behavior of the radial profiles suggests that the NNW edge of the ``tear-drop'' shape is a cold front. The best-fit power-law indices across the edge are $\alpha_1 = 0.32 \pm 0.04$ and $\alpha_2 = 0.94 \pm 0.24$. The best-fit density jump factor is $1.4 \pm 0.2$. The edge radius, obtained from the fit, is $r_{\text{edge}} = 482.3 \pm 18.2$\arcsec, and is shown by the dashed line in Fig.\ \ref{fig:profs} (upper left panel). Due to the asymmetric nature of the cluster emission, the center of origin for the region labeled CF1 was chosen to match the curvature of the feature of interest, rather than centered on the cluster center. Thus, the profile for region CF1 has radial coordinates that are not relative to the cluster center. 

\paragraph{Region CF2} There is a drop in X-ray brightness seen near $\sim$ 225\arcsec. At this radius, the temperature increases from \fe{5.0}{0.6}{0.5} to \fe{6.4}{0.8}{0.7}, while the pressure remains roughly constant across the edge, consistent with the presence of a cold front. The best-fit power-law indices across the edge are $\alpha_1 = 0.38 \pm 0.13$ and $\alpha_2 = 0.79 \pm 0.03$. The best-fit density jump factor is $1.21\pm 0.10 $. The edge radius obtained from the fit is $r_{\text{edge}} = 228.2 \pm 3.4$\arcsec, and is indicated by the dashed line in Fig.\ \ref{fig:profs} (upper right panel). This is consistent with the brightness edge north of the cluster center that is noted in \S \ref{sec:ggm} and Fig.\ \ref{fig:ggm} (cyan arrows). Note that while the bright spot seen just inside the edge (blue arc) of CF2 in Fig.\ \ref{fig:fronts} was not detected as a point source, we define a circular region containing that clump and exclude it from the surface brightness and spectral extractions. This was done in order to ensure proper detection of the cold front in that region.

\paragraph{Region SF1} The radius of curvature for region SF1 was chosen to better fit the observed edge rather than the cluster center. Therefore, while the surface brightness profile of SF1 was fit using the radii from the off-centered coordinates, the profiles for SF1 plotted in Fig.\ \ref{fig:profs} (bottom panel) show the radius in terms of distance from the cluster center. The radial profiles for region SF1 show a drop in brightness, temperature, density, and pressure at a radius of 265\arcsec\ from the cluster center. The temperature drops from \fe{7.6}{1.2}{0.9} to \fe{5.5}{0.6}{0.5} while the pressure decreases by a factor of \fe{1.8}{0.5}{0.4}, consistent with a shock front. The best-fit power-law indices are $\alpha_1 = 0.04\pm 0.6$ and $\alpha_2 = 1.14\pm 0.23$. The density jump factor, obtained from the fit, is $1.31\pm 0.16$. The best-fit edge radius is at a distance of $266\pm 10$\arcsec\ from the cluster center, and is indicated by the dashed line in Fig.\ \ref{fig:profs} (bottom panel). This is consistent with the brightness edge south of the cluster center seen in the GGM filtered image (see \S\ref{sec:ggm} and yellow arrows in Fig.\ \ref{fig:ggm}).

\section{Discussion}

\subsection{Cold Fronts and a Shock }

Figure \ref{fig:fronts} shows the cold fronts and shock front that were identified from the radial profiles discussed above. We have two regions, CF1 and CF2, whose radial profiles exhibit behaviors consistent with cold fronts. The exact geometry of the connection between the two cold fronts in A119 is not perfectly clear. Unfortunately, when examining regions just south of the cluster core, we were unable to confidently identify a cold front in this area. This was mostly due to the inability to disentangle a potential cold front from both the X-ray deficient cavity near the eastern NAT and what could possibly be regions of shock heated gas. However, we propose that the two cold fronts may be connected through a sloshing spiral structure, beginning at large radii in the north with CF1, and spiraling inwards in a clockwise direction to the south, continuing towards the east, and finally NNW to region CF2. This is shown schematically in Fig.\ \ref{fig:fronts} where we indicate with dashed lines how the two cold fronts might be portions of a large-scale sloshing spiral structure. The presence of a sloshing cold front could explain a few of the observed peculiarities of A119. The possible sloshing spiral structure could correspond to the elongated X-ray emission (i.e. the ``teardrop'') seen in the NE. The regions of displaced cool gas seen in the temperature map could have been the result of sloshing induced motions of the ICM. A sloshing spiral could also account for the sudden drop in abundance seen in Fig.\ \ref{fig:globalprof} as the sloshing motion tends to displace the metal-rich gas from the center, bringing it in contact with the metal-poor gas of the surrounding ICM. Our proposed scenario for how a sloshing spiral could have formed in A119 as a result of an off-axis merger is explored further in \S \ref{sec:sims}.

Region SF1 shows compelling evidence of the presence of a shock front located $\sim$250\arcsec\ south of the cluster center. From the Rankine Hugoniot jump conditions, we calculate the Mach number, $M$, of the shock front in terms of the density jump \citep{Landau1959}:
\begin{equation}
        M = \sqrt{\frac{2r}{\gamma + 1 - r (\gamma -1)}}
\end{equation}
where $r = \frac{\rho_1}{\rho_2}$, the subscripts 1 and 2 refer to the density upstream (pre-shock) and downstream (post-shock), respectively, and $\gamma = \frac{5}{3}$ is the adiabatic index assuming ideal gas. Using the density jump obtained from the fit to the radial profile of region SF1, $r = \frac{\rho_1}{\rho_2} = 1.31 \pm 0.16$, we calculate a Mach number of $\mathcal{M} = 1.21 \pm 0.11$. Using the sound speed of the cluster (see \S \ref{sec:totalspec}), we estimate the velocity of the shock front to be $v_{\text{shock}} = c_s\mathcal{M} \simeq 1530 \pm 140$ km/s. A temperature drop from \fe{7.6}{1.2}{0.9} to \fe{5.5}{0.6}{0.5} with a pressure decreases of a factor of \fe{1.8}{0.5}{0.4} is also measured across the shock (see \S\ref{sec:radprof}).

A shock along the southern outskirts of the cluster could have formed as a result of merger activity. This merger shock, when included in the overall picture with our suggested sloshing spiral structure described above, suggests A119 could have experienced recent or ongoing off-axis merger activity along the N-S axis. This scenario is further explored through comparison with simulations of cluster mergers in \S\ref{sec:sims}.

\subsection{Velocity Estimates of Tailed Radio Galaxies \label{sec:vel}}

As discussed in the introduction, the tails of NAT radio sources are bent via ram-pressure, which causes the radio jets to be bent backward. This can arise from high orbital velocities of the host galaxy, bulk motion of the ICM, or some combination of the two. A119 hosts two NAT sources which, notably, have their tails oriented in roughly the same direction, along the NE-SW axis. The similar tail orientations, along with both tails being aligned with the overall NE-SW elongation of the X-ray emission, suggest a possible merger origin for the two NATs. To determine whether this scenario is feasible, we calculate the velocity of the host galaxies relative to the ICM that would be required in order to produce the observed radio jet/lobe bending.

To estimate the velocities of the two NAT sources, we follow the methods of \cite{Douglass2008, Douglass2011} and \cite{Paterno-Mahler2013}. If we assume that the radio lobes are bent only due to the ram pressure from the surrounding ICM, and assuming pressure balance between the lobes and the ICM, we can apply Euler's equation
\begin{equation}
    \frac{\rho_r v^2_r}{r_c} = \frac{\rho_{\text{\tiny ICM}} v^2_g}{r_r}
    \label{eq:euler}
\end{equation}
where $\rho_r$ is the mass density of the lobe, $r_r$ is the radius of the lobe, $v_r$ is the velocity of the plasma in the lobe, $r_c$ is the radius of curvature, $\rho_{\text{\tiny ICM}}$ is the mass density of the ICM, and $v_g$ is the velocity of the galaxy relative to the ICM. For the mass density of the ICM, we take $\rho_{\text{\tiny ICM}} = 1.92 n_e \mu m_p$, where $\mu$ is the mean atomic mass of the cluster, taken to be 0.6, and $n_e$ is the gas density found from the density profiles. If we assume in situ particle acceleration in the lobes \citep{ODonoghue1993}, then the condition used to determine the internal density required to produce the observed luminosity of the lobes is 
\begin{equation}
    L_{rad} = \frac{1}{2} \epsilon \pi r^2_r \rho_r v^3_r
    \label{eq:Lrad1}
\end{equation}
where $\epsilon$ is the radiative efficiency and has typical values between $0.001$ and $0.1$ \citep{Birzan2004}. The radiative efficiency measures the conversion of kinetic energy into the observed radiation. It is defined as the ratio of total radio luminosity, $L_{rad}$, to kinetic luminosity, $L_{kin} = 4PV/t$ (for relativistic jets; \cite{Churazov2001}). To calculate the kinetic luminosity, we can estimate the ICM pressure from our pressure profiles. The volume of the NAT region was calculated assuming an elliptical geometry. We assume a typical AGN repetition rate of $t = 5\times10^{7}$ yr \citep{Birzan2004}. The total radio luminosity is calculated by integrating the flux between $\nu_1 = 10^7$ Hz and $\nu_2 = 10^{11}$ Hz
\begin{equation}
    L_{rad} = 4\pi D^2_L S_{\nu_0} \int^{\nu_2}_{\nu_1} \bigg(\frac{\nu}{\nu_0}\bigg)^{-\alpha} d\nu
    \label{eq:Lrad2}
\end{equation}
For both NAT sources, we use a spectral index of $\alpha = 0.7$, calculated for these sources in \cite{Feretti1999} from 1.4 GHz, 4.9 GHz, and 8.4 GHz VLA observations. Reference fluxes and frequencies are taken from the VLA FIRST catalogs. 

Combining the luminosity condition of Eq. \ref{eq:Lrad1} with Euler's equation from Eq. \ref{eq:euler}, the velocity of the host galaxy relative to the ICM that would be required bend the lobes is 
\begin{equation}
    v_g = \bigg(\frac{2 L_{rad}}{\epsilon \pi \picm v_r r_c r_r}\bigg)^{1/2}
    \label{eq:vg}
\end{equation}

\begin{figure*}
\begin{center}$
\begin{array}{cc}
\includegraphics[width = 0.6 \textwidth]{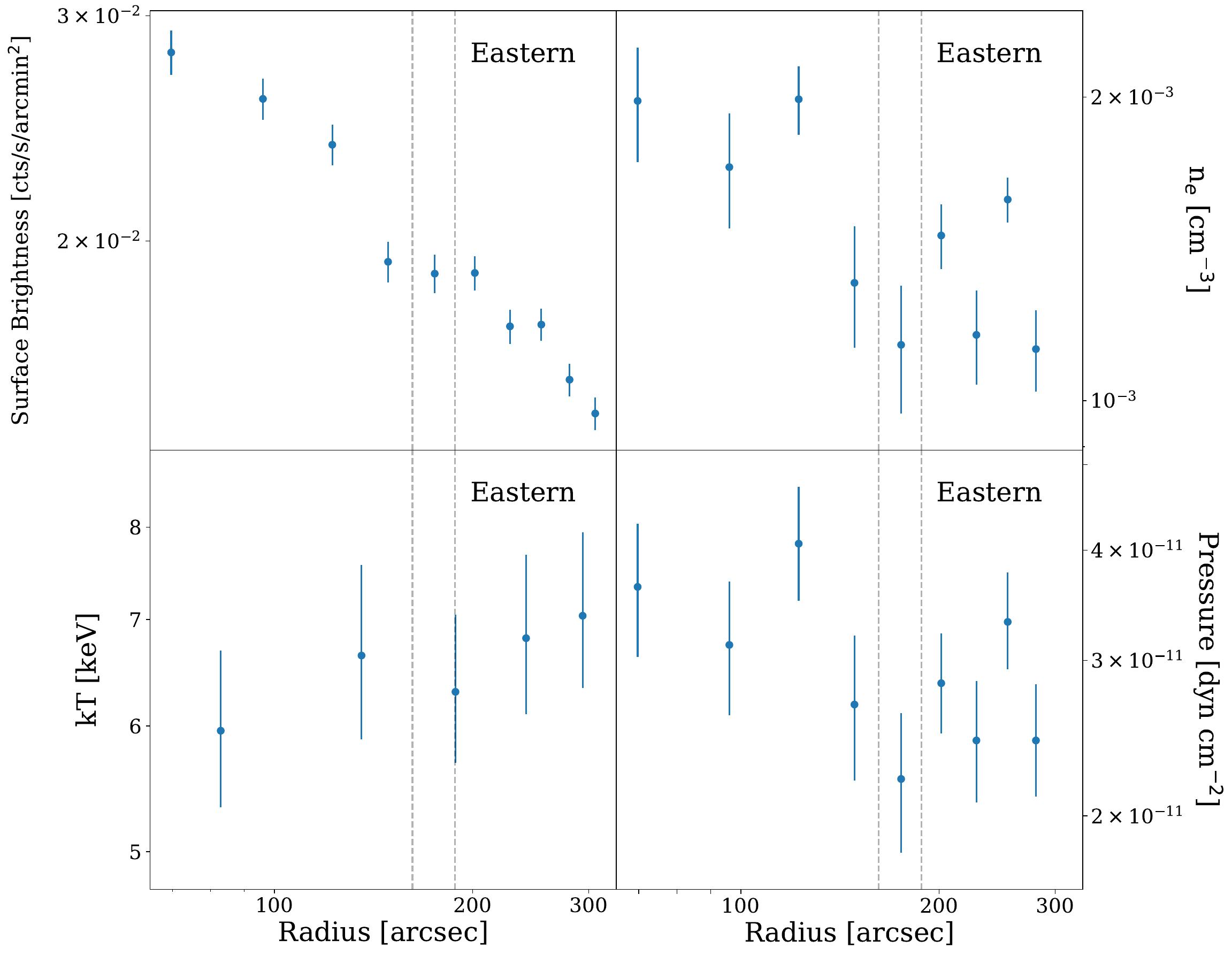} & 
\includegraphics[width = 0.4 \textwidth]{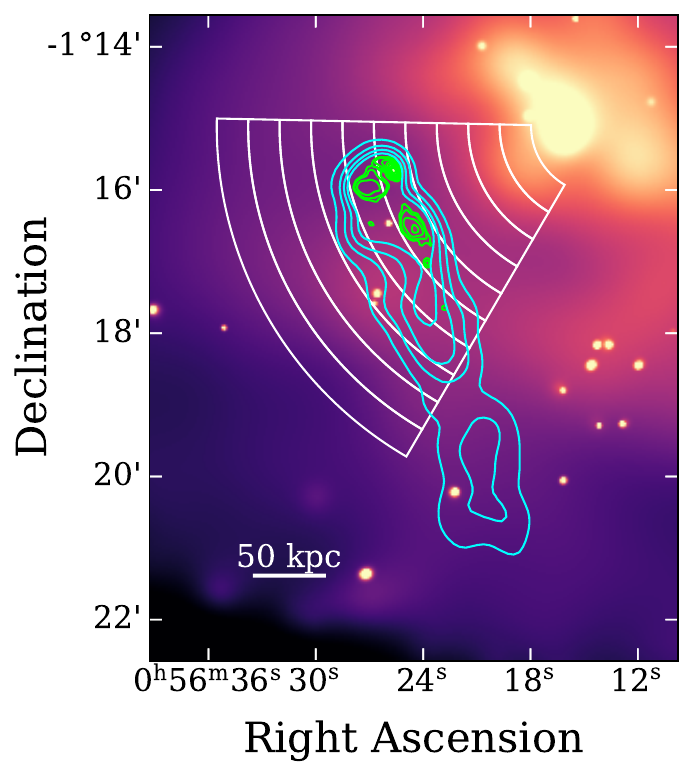}
\end{array}$
\caption{Radial profile across the eastern NAT source used to calculate the velocities in \S \ref{sec:vel}. Dashed lines denote the radial extent of the bin, containing the NAT host galaxy, from which the pressure and density are estimated.\label{fig:nat15}}
\end{center}
\end{figure*}

\begin{figure*}
\begin{center}$
\begin{array}{cc}
\includegraphics[width = 0.6 \textwidth]{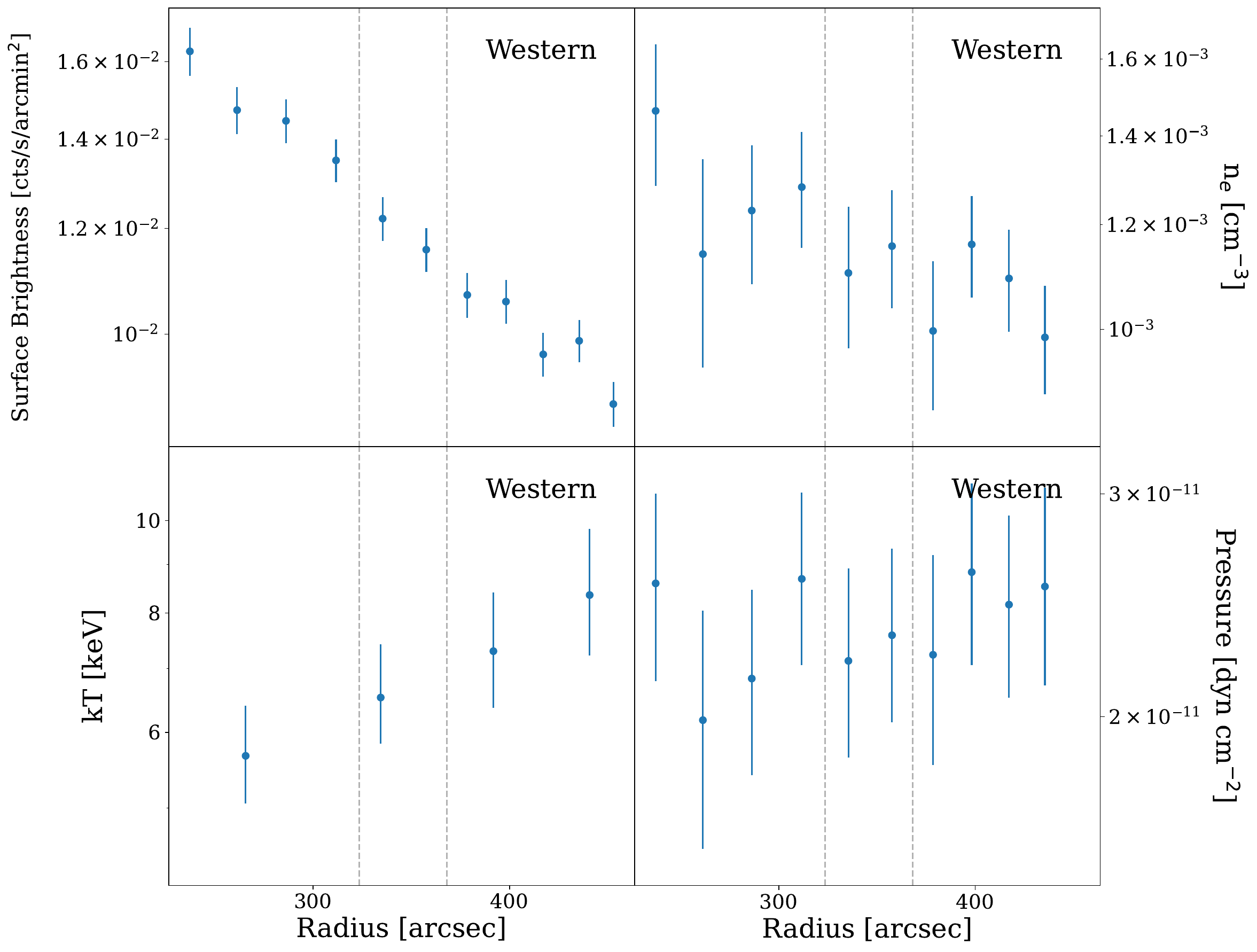} & 
\includegraphics[width = 0.4 \textwidth]{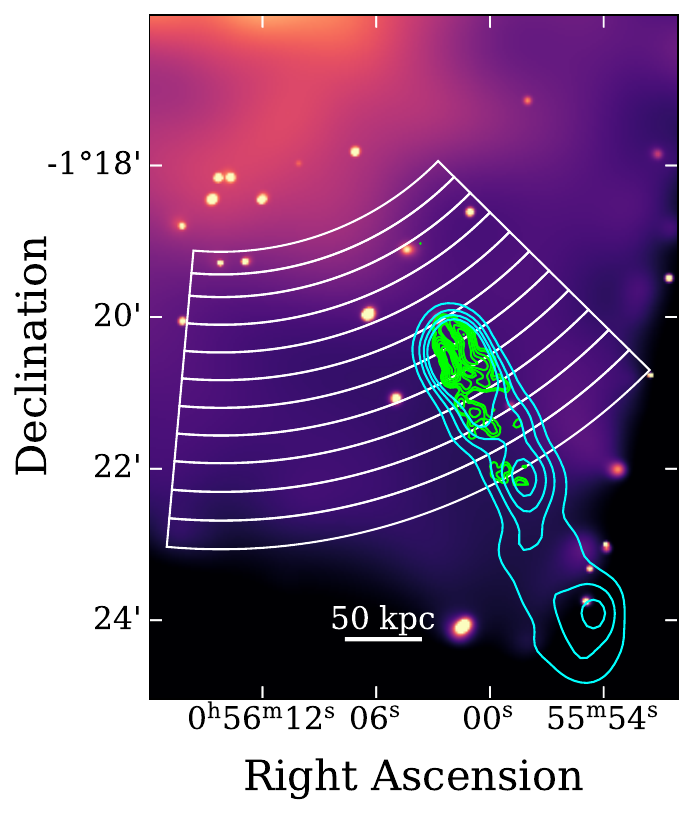}
\end{array}$
\caption{Radial profile across the western NAT source used to calculate the velocities in \S \ref{sec:vel}. Dashed lines denote the radial extent of the bins, containing the NAT host galaxy, from which the pressure and density are estimated. \label{fig:nat16}}
\end{center}
\end{figure*}

For the velocity of the plasma in the radio lobes, $v_r$, we define upper and lower limits of 0.08c to 0.2c \citep{ONeill2019b, TerniDeGregory2017}. To calculate $\epsilon$ and $\picm$, we need to determine estimates of the pressure and density of the ICM surrounding the NAT sources. For the eastern NAT, we estimate the surrounding ICM pressure and density from the radial profile shown in Fig.\ \ref{fig:nat15}, between radii of $\sim 162-188\arcsec$. The pressure and density of the ICM surrounding the western NAT is estimated from the radial profile shown in Fig.\ \ref{fig:nat16} between radii of $\sim 321-365\arcsec$. Note that we assume a plane of sky orientation for the two NAT sources and therefore, due to projection effects, the actual distances from the NAT sources to the cluster center may be different than the assumed distances.

For each of the NAT sources, we use Eq.\ \ref{eq:Lrad2} to calculate the total radio luminosity. From the total radio and kinetic luminosities, we calculate the efficiency ($\epsilon$). These are then used in Eq.\ \ref{eq:vg} to determine the velocity of the host galaxy relative to the ICM that would be required to produce the observed bending of the NAT. The results for the individual NAT sources are summarized below. 

\paragraph{Eastern NAT (0053-015)} The total radio luminosity of 0053-015 was calculated to be $3.5\times10^{40}$ erg s$^{-1}$ using a reference flux of $47.5$ mJy at a frequency of 1435 MHz. To calculate the kinetic luminosity, we need to determine the pressure of the overlying ICM around the NAT, as well as an estimate of the volume of the NAT. The pressure of the surrounding ICM at the radius of the eastern NAT is found to be $P = 2.2\times 10^{-11}$ dyne cm$^{-2}$. The volume of the NAT region was calculated assuming an ellipsoidal geometry with $a = 16.5$ kpc and $b = 13.6$ kpc. We calculate a kinetic luminosity of $2.1\times10^{43}$ erg s$^{-1}$. From the radio and kinetic luminosities, we find an efficiency of $\epsilon = 0.0016$.
The mass density of the ICM was calculated to be $\rho_{\text{\tiny ICM}} = 2.1\times10^{-27}$ g cm$^{-3}$, where the gas density of the surrounding ICM at the radius of the NAT is $n_e = 1.1\times 10^{-3}$ cm$^{-3}$. The radius of the lobe, measured from the 1.4 GHz maps provided in \cite{Feretti1999}, was found to be $r_r = 6.6$ kpc, measured from the eastern lobe which displays a greater degree of bending. The radius of curvature was found to be $r_c = 19.8$ kpc, again taken from the curvature of the eastern lobe and measured from the 1.4 GHz map. It should be noted that the bending in the eastern-most jet occurs near the edge of the visible stellar halo of the galaxy. This could suggest that the galaxy has retained some of its interstellar medium (ISM) which could be protecting the jet until it reaches the ISM/ICM interface, at which point the jet is swept back. Thus our measurement of the radius of curvature for this source should be considered an upper limit. 

From the values discussed above, we find that the relative velocity, in the plane of the sky, between the host galaxy and the ICM (Eq.\ \ref{eq:vg}) is between 290 km s$^{-1}$ and 450 km s$^{-1}$, consistent with values previously seen in the sloshing-induced bending of radio lobes \citep{Paterno-Mahler2013}. Additionally, \cite{Feretti1999} gives the peculiar radial velocity, along the line of sight, relative to the average cluster velocity as $|\Delta v| = 1697$ km s$^{-1}$ indicating a significant velocity component from the galaxy motion itself.

\paragraph{Western NAT (0053-016)} The total radio luminosity of 0053-016 was calculated as $2.3\times 10^{41}$ erg s$^{-1}$ using a reference flux of 314.2 mJy at a frequency of 1435 MHz. The kinetic luminosity was found to be $1.1\times10^{44}$ erg s$^{-1}$, where the pressure of the surrounding ICM at the radius of the western NAT is $P= 2.3\times 10^{-11}$ dyne cm$^{-2}$, and the volume was calculated assuming an ellipsoidal geometry with $a = 32.2$ kpc and $b = 22.3$ kpc. From the total radio and kinetic luminosities, we find an efficiency of $\epsilon = 0.0020$. 

From the density profile, we estimate the gas density of the surrounding ICM at the radius of the NAT to be $n_e = 1.1\times 10^{-3}$ cm$^{-3}$, giving an ICM mass density of $\rho_{\text{\tiny ICM}} = 2.1\times 10^{-27}$ g cm$^{-3}$. From the 1.4 GHz maps of \cite{Feretti1999}, we measure $r_r = 1.9$ kpc and $r_c = 12.3$ kpc. 

Using the values above in Eq.\ \ref{eq:vg}, we calculate the velocity, in the plane of the sky, of the host galaxy relative to the ICM to be between 1572 km s$^{-1}$ and 2486 km s$^{-1}$. These high relative velocities suggest that the bending of the western NAT's lobes is likely not solely induced from ICM sloshing motions. Additionally, \cite{Feretti1999} gives the peculiar radial velocity, along the line of sight, relative to the average cluster velocity as $|\Delta v| = 415$ km s$^{-1}$, indicating the predominant galaxy motion is in the plane of the sky. The western NAT appears to lie at a much further radius from the cluster core and we would expect higher galaxy orbital velocities. Therefore, there is likely some combination of both ICM bulk flow and galaxy orbital motion that contribute to the shaping of the western NAT.

\subsection{Optical Substructure in A119}

\begin{figure}
    \plotone{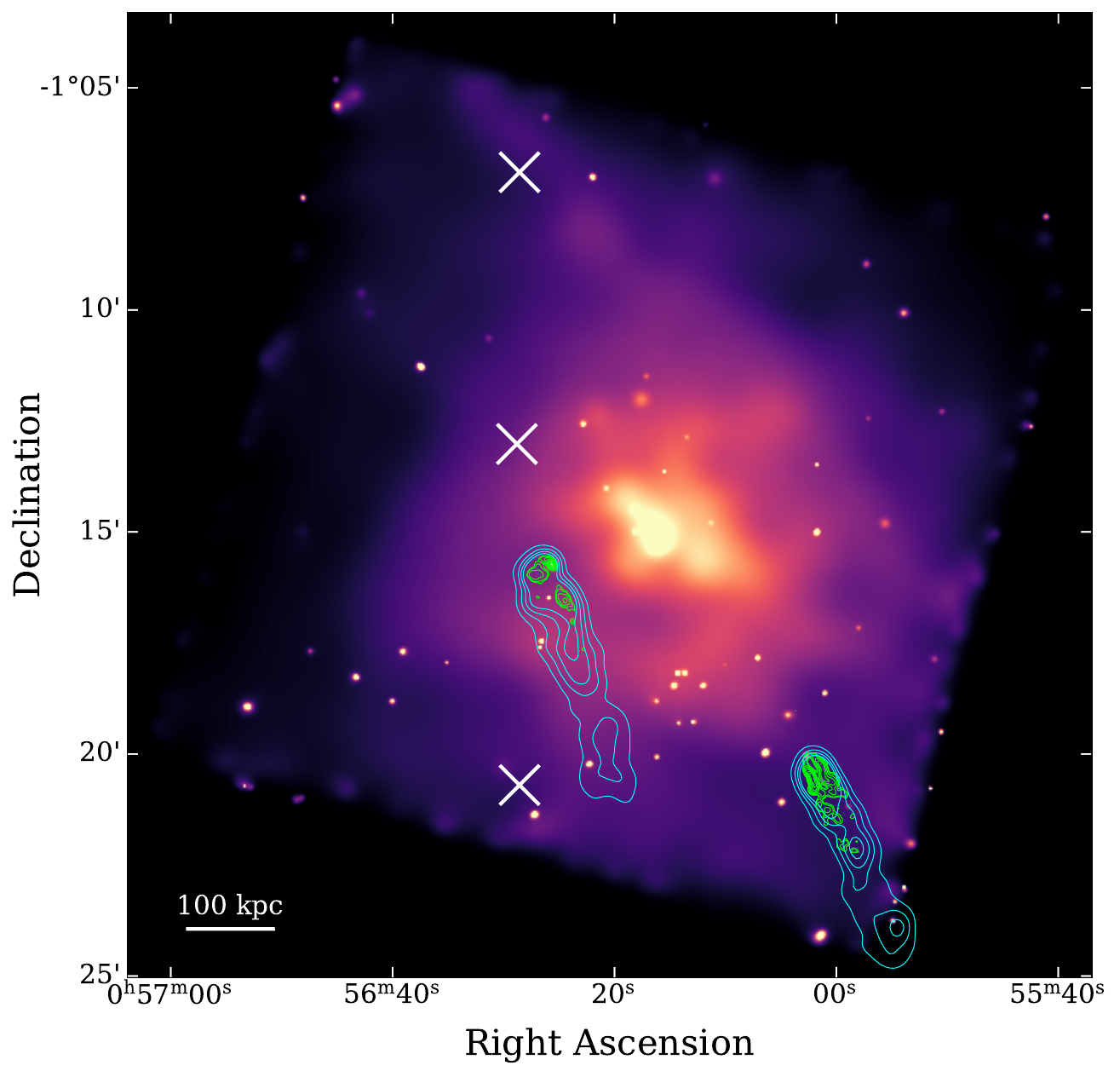}
    \caption{Adaptively smoothed X-ray image showing locations of three optical subclusters reported in previous studies. \citep{Fabricant1993, Kriessler1997,Ramella2007, Lee2016}.}
    \label{fig:subgrps} 
\end{figure}

Figure \ref{fig:subgrps} shows the locations of three subclusters identified in previous optical studies \citep{Fabricant1993, Kriessler1997,Ramella2007, Lee2016}. The substructures are oriented in a nearly straight line along the N-S axis, providing further evidence of a merger occurring along this direction. In addition to the three subclusters shown, \cite{Tian2012} identify possible substructure located nearly directly foreground or background to the core. They suggest this is indicative of merger activity in the cluster core that is occurring along the line of sight.

Along with the alignment of optical substructures, the overall galaxy distribution of A119 has also been shown to follow the NE-SW trend. \cite{Lee2016} found that within a 0.64 Mpc radius of the cluster center the galaxy distribution of cluster members is elongated to the NE. They further show that the large-scale distribution of galaxies with radial velocities similar to that of cluster galaxies form a filamentary structure that also extends to the N-NE. This filament appears to extend nearly 8 Mpc north, connecting A119 to another galaxy cluster, A116 ($z=0.066$) (see Fig. 11 of \cite{Lee2016}). 

It is interesting to note that, if A119 is connected to A116 through a large-scale filament, the orientation of the two NAT tails are also aligned along this same axis. This has been seen previously in the alignment of jets from wide-angle tail (WAT) radio sources with the orientation of the supercluser in which the cluster belong. \cite{Novikov1999} showed 98\% confidence in the alignment of WAT jets and supercluster axes and they suggest this could be the result of remnant drainage of matter along the large-scale filamentary structures connecting the neighboring clusters. While the \cite{Novikov1999} study focused on WAT rather than NAT sources, intriguingly, we observe a similar occurrence in A119 of the two NAT tails having general alignment with the large scale filamentary structure seen in \cite{Lee2016}.

\subsection{Comparison With Simulations\label{sec:sims}}

Cold fronts and shock fronts are produced by merger activity. Sloshing cold fronts, in particular, are produced by an encounter of a massive cluster with a much smaller subcluster that passes by with a non-negligible impact parameter \citep{Ascasibar2006}. To further refine the merger scenario for A119, we examined the idealized binary merger simulations from the Galaxy Cluster Merger Catalog\footnote{\url{https://gcmc.hub.yt}} (\cite{ZuHone2018}; hereafter GCMG) which are of this type. Our observation appears to be qualitatively reproduced by a simulation of a 1:10 mass-ratio merger with an initial impact parameter of $\sim$1~Mpc, a simulation originally presented in \cite{ZuHone2011a}. While the simulation results provide useful comparisons to our observations, it should be noted that this simulation was not originally designed to reproduce the features in A119.

In this simulation, the subcluster initially passes by the main cluster with a closest approach of a $\sim$few hundred kpc. This first passage initiates the sloshing motions. Because the subcluster is on a bound orbit with the main cluster, it returns to the main cluster core $\sim 3.7$ Gyr later, narrowly missing the core this time, and drives a shock front. This shock front coincides roughly with the semicircular southern shock (the red line shown in the cartoon schematic of Fig.\ \ref{fig:fronts}) in A119. 

\begin{figure*}
\gridline{\fig{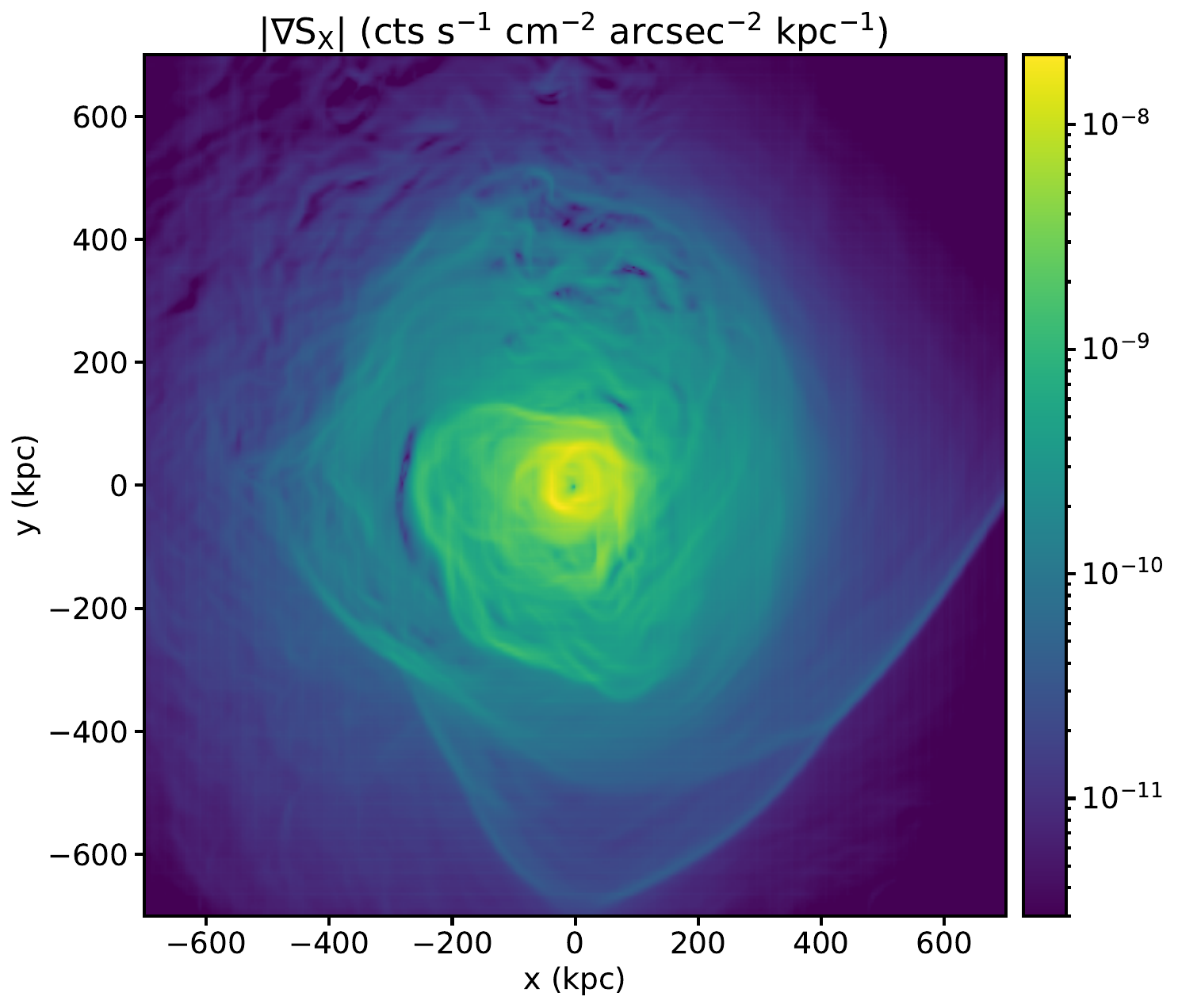}{0.5\textwidth}{}
          \fig{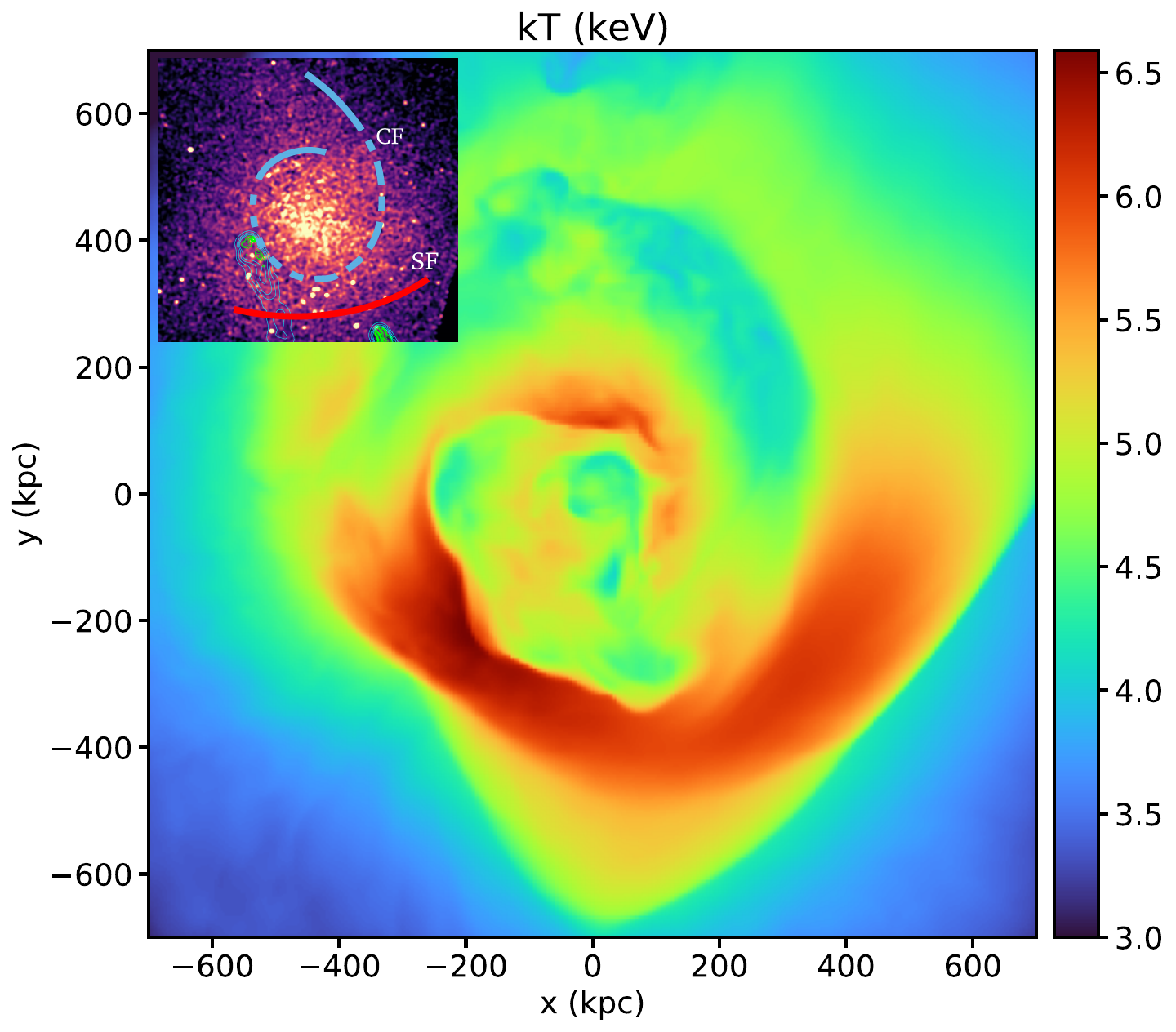}{0.5\textwidth}{}}
    \caption{A cluster merger simulation which qualitatively matches our preferred scenario for A119. \emph{Left}: Gradient-filtered X-ray surface brightness (0.5-7.0 keV band). \emph{Right}: Projected spectroscopic-like \citep{Mazzotta2004} gas temperature. We include an inset of the cartoon schematic from Fig.\ref{fig:fronts} to aid in comparison of the features observed in A119 with the simulation results.\label{fig:sims}}
\end{figure*}

The particular epoch of this simulation that best matched our observations is $\sim 0.2$ Gyr after this second core passage, which has a shock front surrounding the sloshing cold fronts\footnote{\url{http://gcmc.hub.yt/fiducial/1to10_b1/0255.html}}. This is shown in Figure \ref{fig:sims}, which shows a gradient-filtered image of the X-ray surface brightness and projected temperature from the simulation. It has been rotated so that the positions of the cold and shock fronts approximately match those in A119, and that the merger trajectory occurs in the NE-SW direction, as in our preferred merger scenario. The relatively close match of this simulation to the observed features of A119 indicates that a small subcluster approached from the SW, passed by the main cluster from the south, traveled to the NNW before turning around and passing the main cluster a second time from the N. Currently, the whereabouts of the subcluster is unknown, but depending on the exact line-of-sight angle, it may be co-aligned with the main cluster on the sky or may be located to the SW. A simulation more tailored to match A119 would be necessary to resolve this question.

\section{Summary \& Conclusions}

We have presented new \chandra observations of the dynamically complex cluster Abell 119. The X-ray emission from the ICM is asymmetric with an elongation to the NE, resulting in a tear-drop like shape clearly seen in the \chandra image. Residual images show evidence of underlying structure in the ICM that is explored in depth using radial profiles of the X-ray brightness, temperature, density and pressure.

From the extracted radial profiles we identify two cold fronts which we propose could be connected to form a sloshing spiral structure (see the cartoon schematic in Fig.\ \ref{fig:fronts}). Spectral maps show regions of cool gas along the possible spiral, consistent with the cooler gas of the cluster core being pulled out to larger radii via sloshing. If this is a sloshing cold front, A119 could also provide useful insight into the role of off-axis mergers in the formation of NCC clusters.

A surprising discovery was the presence of a shock front located $\sim 250$\arcsec\ from the core and just outside of the potential sloshing cold front. From the temperature profile we measure a temperature drop from \fe{7.6}{1.2}{0.9} to \fe{5.5}{0.6}{0.5} across the shock. We calculate a Mach number for the shock of $\mathcal{M}= 1.21 \pm 0.11$, corresponding to a velocity of $v_{\text{shock}} = 1530 \pm 140$ km/s, when using the sound speed of the cluster calculated in \S \ref{sec:totalspec}.

Another remarkable feature of this cluster is the orientation of the two NAT sources with each other as well as the asymmetric X-ray emission. Both NAT tails are oriented nearly parallel to each other despite the fact that the jets are actually leaving their hosts in very different directions. In addition to this, both tails are aligned with the NE-SW elongation of the X-ray emission. We examine this further by using our results from the radial profiles to calculate the velocities of the NAT host galaxies relative to the ICM. Velocities of the eastern NAT are found to be consistent with values expected from sloshing-induced bending. However, \cite{Feretti1999} reports a significant line of sight velocity suggesting there is some contribution to the bending of the tail from the galaxy motion itself. The western NAT was found to have fairly high relative velocities. This, along with its distance from the cluster center, suggest there is some combination of ICM bulk flow with the host galaxy infall motions that is working to produce the observed bending.

Following the NE-SW alignment trend of the X-ray emission and NAT tails, \cite{Lee2016} showed that the optical distribution of cluster members is likewise elongated to the north and the presence of optical substructures oriented along the N-S direction. Beyond this, there is evidence of cluster galaxies forming a filamentary structure which extends nearly 8 Mpc to the N-NE which appears to connects A119 to A116. The alignment of the two NAT sources with this filament could also be suggestive of cluster winds along the supercluster axis \citep{Novikov1999}.

Comparison of our observations with the simulations of \cite{Zuhone2011} helped to form a clearer picture of a possible merger scenario which we propose could explain both the cold fronts and the shock in A119. We find that the sloshing cold front produced as a result of the first passage of a much smaller perturbing subcluster approaching from the SW. The simulations also suggest that the shock front in the south could then be the result of a secondary passage several Gyr later, with the subcluster now coming in from the north. While these simulations were not originally designed to reproduce the exact dynamics of A119, they provide a very useful visual comparison to our observations. 

A119 has proven to be a fairly complex system, playing host to two NAT sources, a potential sloshing spiral, a merger shock, and possible connection with a neighboring cluster through large-scale filamentary structures. Untangling these features and investigating how they are connected to or interacting with the cluster is important to further our understanding of cluster formation and evolution. Our results show alignment of the elongated X-ray emission, the optical substructures, and the flow directions of the radio jets/lobes of both NATs. This, with the comparison to simulations, all supports our picture of recent or on-going merger activity occurring in the NE-SW direction. 

\acknowledgements

We thank the anonymous referee for their helpful comments that greatly helped improve this manuscript. CBW was supported by the National Aeronautics and Space Administration through SAO Chandra Cycle 21 Award GO0-21123X, XMM-Newton Cycle 17 Award 80NSSC19K0743, and STScI HST Cycle 27 Award GO-15994. CBW was also supported by Massachusetts Space Grant Consortium Awards 580379, 633061, 699842, 742790, 793224, and 858617. Additional support for this work was partially provided by the Chandra X-ray Center through NASA contract NAS8-03060, the Smithsonian Institution, and by the Chandra X-ray Observatory grant GO9-20112X.

\bibliography{main.bib}

\end{document}